\documentclass[11pt,a4paper]{article}

\usepackage[utf8]{inputenc}
\usepackage[T1]{fontenc}
\usepackage{lmodern}

\usepackage{amsmath,amssymb,amsthm}
\usepackage{graphicx}
\usepackage{booktabs}
\usepackage{hyperref}
\usepackage{geometry}

\title{
    Probabilistic Cellular Automata: between deterministic Wolfram's rules 23, 77, 178 and 232
}

\author{
    Francisco  J. Mu\~noz$^{1}$, Juan Carlos Nu\~no$^{1}$ \\[0.3cm]
    $^{1}$Departamento de Matem\'atica Aplicada \\
    Universidad Polit\'ecnica de Madrid \\
     28040, Madrid, Spain \\[0.2cm]
    \texttt{fcomunoz@alumnos.upm.es, juancarlos.nuno@upm.es}
}

\date{}

\begin{document}

\maketitle

% -------------------------------------------------
% ABSTRACT
% -------------------------------------------------
\begin{abstract}
We study one‑dimensional binary Probabilistic Cellular Automaton (PCA) that interpolate between Wolfram’s classical rules 23, 77, 178 and 232.  These rules are the only ones that satisfy two criteria: (i) in the case of a majority in the neighborhood states, the central site takes either the majority state or the opposite and (ii) if the neighborhood states are tied, the central site either changes its current state or keeps it. The PCA is defined by two Bernoulli random variables with parameters $p,r\in [0,1]$, and we analytically solve small-size cases by using a Markov process formulation. We derive analytical expressions for the probability of asymptotically reaching each possible global configuration as a function of $p$ and $r$, for all initial states. We show that for $0 < p,r < 1$, the asymptotic probability distributions of achieving any of the states for the PCA are independent of the initial conditions. This contrasts with the behavior of the deterministic Wolfram's rules 23 ($p=0,r=0$), 77 ($p=1,r=0$), 178 ($p=0,r=1$) and 232 ($p=1,r=1$), for which additional asymptotic states can occur, in particular periodic configurations. Finally, we discuss applying this kind of PCA to describe opinion dynamics involving hesitant agents.
\end{abstract}

\vspace{0.3cm}


% Main text
\section{Introduction}

Cellular automata (CA) are prototypical deterministic discrete dynamical systems that exhibit a wide variety of behaviors \cite{Wolfram1983,Wolfram}. Elementary Wolfram CA are one-dimensional arrays of cells with two possible states, evolving strictly based on the state of each cell and its nearest neighbors (radius 1) \cite{Wolfram1983,Wolfram}. Accordingly, there are $2^{2^3}$ possible rules, each assigning the next state of the central cell to every possible triplet configuration. These rules can be encoded as binary numbers obtained from this mapping, as well as by their corresponding decimal representations. Wolfram’s classical CA are deterministic: their evolution is fully determined by the initial conditions and the update rule. In the next section, we summarize the main properties of the four Wolfram rules: 23, 77, 178, and 232. 

\begin{figure}[htbp]
  \centering
  \includegraphics[width=0.4\textwidth]{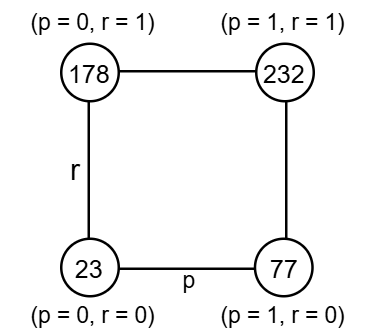}
  \caption{Probabilistic parameter space defined by $p,r \in [0,1]$. The four vertices correspond to the deterministic Wolfram's rules: 23 $(p=0,r=0)$, 77 $(p=1,r=0)$, 178 $(p=0,r=1)$ and 232 $(p=1,r=1)$. The horizontal edges describe rules with non-standard ($r=0$) and standard ($r=1$) contagion in case of majority. Vertical edges represent rules that, in case of tie, keep the state of the central cell ($p=0$) or change this state ($p=1$). In the interior of the square, $0<p,r<1$, the PCA exhibits stochastic dynamics that depends on the probabilistic parameters.}
  \label{prsimplex}
\end{figure}

In contrast, Probabilistic Cellular Automata (PCA) are stochastic, as their update rules involve randomness. Consequently, their dynamics are not uniquely determined by the initial conditions, but instead correspond to realizations of an underlying stochastic process. In this work, we introduce a PCA that stochastically interpolates between the deterministic rules mentioned above. The model depends on two probabilistic parameters, $p$ and $r$, that vary over the square $[0,1] \times [0,1]$ (see Fig.~\ref{prsimplex}). The asymptotic dynamics are analyzed using a Markov process. In Section \ref{seclowsize}, this approach is applied to small system sizes, allowing rigorous comparison with the deterministic cases at the vertices of the parameter square. We assume periodic (cyclic) boundary conditions and synchronous updating.

This type of probabilistic cellular automaton is used to study opinion dynamics \cite{Castellano2009, SanMiguel2005, Munoz2025} in systems where agents (sites or cells) experience hesitation. In particular, the model considers the effect of uncertainty in two complementary situations: (i) majority and (ii) counter-majority interactions. In the first case, when an agent’s neighborhood results in a tie, the update rule depends on the stochastic parameter $p$, which represents the probability of maintaining the current state. In the second case, the agent hesitates in responding against the majority opinion, and this behavior is quantified by a second stochastic parameter, $r$. Taken together, these two parameters enable a comprehensive study of the influence of hesitation on social dynamics.

In this paper, we study low-dimensional cyclic cellular automata with the main goal of obtaining an analytical solution to the model and rigorously determining the influence of the stochastic parameters on the dynamics. Specifically, the characterization of the discontinuities that arise at the boundaries of the probabilistic region. Extending these results to larger system sizes is challenging due to the exponential growth of the transition matrix and the resulting increase in the complexity of its algebraic analysis.

The remainder of the paper is organized as follows. In the next section, we present the deterministic rules that delimit the probabilistic regime. In Section \ref{PCA}, we define the probabilistic cellular automata, which is solved exactly in Section \ref{seclowsize} for system sizes $L=2$ and $L=3$. Finally, Section \ref{concluding} contains the concluding remarks.

\section{Deterministic rules}

The vertices of the square defined by $p,r \in [0,1]$ correspond to the deterministic Wolfram rules, briefly described here. These rules are defined for cyclic one-dimensional CA with radius 1, i.e. each site is influenced exclusively by its nearest neighbors,  of size $L$ and synchronous updating. The four rules described below belong to Wolfram’s Class II \cite{Wolfram1983}, as they exhibit stable and periodic behavior for almost all initial conditions. 

%Figure \ref{Combinedrules} illustrates their typical dynamics starting from random initial configurations with density 0.5 (i.e., an equal number of 0s and 1s randomly distributed)

\subsection{Rule 23.} This rule occurs when $p=0$ and $r=0$, which means that, in case of majority, the central site takes the opposite state of the two neighbors (non-standard contagion \cite{galam2004}) and, in case of tie, the central site changes its current state (see Table \ref{tab4reglas}).  For $x^i\in \{0,1\}$, $i=1,2,\ldots,L$, the rule map \cite{Fuks23,GarciaMorales2012}, $x^{i}_{t+1} = f_{23}(x^{i-1}_t, x^{i}_t, x^{i+1}_t)$, is given by: 
{
\begin{equation}\label{r23}
f_{23}(x^{i-1}, x^{i}, x^{i+1})  = 2 x^{i+1} x^{i} x^{i-1} - x^{i+1} x^{i}- x^{i+1} x^{i-1} - x^{i} x^{i-1} + 1 
\end{equation}
}

%\begin{figure}[htbp]
%  \centering
 % \includegraphics[width=0.70\textwidth]{Combinedrules.png}
 % \caption{Typical dynamics of the cyclic cellular automata considered in this paper, starting from a random initial configuration with density 0.5, with length $L = 10$, white cells correspond to 0 and black cells to 1: a) Rule 23, b) Rule 77, c) Rule 178, and d) Rule 232 (time evolves from top to bottom).}
  %\label{Combinedrules}
%\end{figure}

%\begin{table}
 %   \caption{Table captions should be placed above the tables.}\label{tab23}
 %   \begin{tabular}{lcccccccc}
  % \hline
   %     \textbf{Configuration}  & 1\textbf{1}1  & 1\textbf{1}0  & 1\textbf{0}1  & 1\textbf{0}0  & 0\textbf{1}1  & 0\textbf{1}0  & 0\textbf{0}1  & 0\textbf{0}0 \\ 
%\hline
   %     \textbf{New State}  & \textbf{0}  & \textbf{0}  & \textbf{0}  & \textbf{1}  & \textbf{0}  & \textbf{1}  & \textbf{1}  & \textbf{1} \\ 
%\hline
 %   \end{tabular}
%\end{table}

\subsection{Rule 77.} As occurs with rule 23, this rule applies a non-standard contagion. However, contrary to rule 23, in case of tie, the central site keeps its current state (see Table \ref{tab4reglas}). For $x^i\in \{0,1\}$,  $i=1,2, \ldots,L$, the map of this rule $x^{i}_{t+1} =f_{77}(x^{i-1}_t, x^{i}_t, x^{i+1}_t)$, is: 
{
\begin{equation}\label{r77}
f_{77}(x^{i-1}, x^{i}, x^{i+1})=  -2 x^{i+1} x^{i}  x^{i-1} + x^{i+1} x^{i}- x^{i+1}+ x^{i+1} x^{i-1}+ x^{i} x^{i-1} - x^{i-1} + 1
\end{equation}
}

%\begin{table}
%    \caption{Table captions should be placed above the tables.}\label{tab77}
  %  \begin{tabular}{lcccccccc}
 %\hline
  %      \textbf{Configuration}  & 1\textbf{1}1  & 1\textbf{1}0  & 1\textbf{0}1  & 1\textbf{0}0  & 0\textbf{1}1  & 0\textbf{1}0  & 0\textbf{0}1  & 0\textbf{0}0 \\ 
 %\hline
  %     \textbf{New State}  & \textbf{0}  & \textbf{1}  & \textbf{0}  & \textbf{0}  & \textbf{1}  & \textbf{1}  & \textbf{0}  & \textbf{1} \\ 
 % \hline
  %  \end{tabular}
%\end{table}

\subsection{Rule 178.} This rule assumes that when the states of neighbors coincide, the central site takes the state of this majority (standard contagion). In case of tie, the central state always changes (see Table \ref{tab4reglas}). For $x^i\in \{0,1\}$, $i=1,2, \ldots,L$, the updating equation for this rule, $x^{i}_{t+1} = f_{178}(x^{i-1}_t, x^{i}_t, x^{i+1}_t)$, is given by:
{
\begin{equation}\label{r178}
 f_{178}(x^{i-1}, x^{i}, x^{i+1})=2 x^{i+1}x^{i} x^{i-1} - x^{i+1} x^{i} + x^{i+1} - x^{i+1} x^{i-1} - x^{i} x^{i-1} + x^{i-1}
\end{equation}
}

%\begin{table}
 %  \caption{Table captions should be placed above the tables.}\label{tab178}
  %  \begin{tabular}{lcccccccc}
   %      \hline
    %    \textbf{Configuration}  & 1\textbf{1}1  & 1\textbf{1}0  & 1\textbf{0}1  & 1\textbf{0}0  & 0\textbf{1}1  & 0\textbf{1}0  & 0\textbf{0}1  & 0\textbf{0}0 \\ 
%   \hline
   %     \textbf{New State}  & \textbf{1}  & \textbf{0}  & \textbf{1}  & \textbf{1}  & \textbf{0}  & \textbf{0}  & \textbf{1}  & \textbf{0} \\ 
 %\hline
  %  \end{tabular}
%\end{table}

\subsection{Rule 232.} This rule assumes that in case of tie in the neighbor's states, the current central state does not change. When majority occurs in the neighborhood, the central site takes the state of the majority (standard contagion). The transition table of this rule is given in Table \ref{tab4reglas}. For $x^i\in \{0,1\}$, $i=1,2,\ldots,L$, the corresponding rule map, $x^{i}_{t+1} =f_{232}(x^{i-1}_t, x^{i}_t, x^{i+1}_t)$, is:
{
\begin{equation}\label{r232}
f_{232}(x^{i-1}, x^{i}, x^{i+1})=-2 x^{i+1} x^{i} x^{i-1} + x^{i+1} x^{i} + x^{i+1} x^{i-1} + x^{i} x^{i-1}
\end{equation}
}

%Contrary to rule 178, this rule can yield asymptotic patterns where all cells are in the same state, either 0 or 1. Nevertheless, the final state depends on the initial configuration. For example, for balance initial configurations, the population is arranged asymptotically in clusters of different sizes.  

\begin{table}
\begin{center}
\caption{Deterministic rules at the vertices of the parameter space.} \label{tab4reglas}
\begin{tabular}{lcccccccc}
\hline
\textbf{Configuration} $(x^{i-1}_t, x^{i}_t, x^{i+1}_t)$ & 1{1}1 & 1{1}0 & 1{0}1 & 1{0}0 & 0{1}1 & 0{1}0 & 0{0}1 & 0{0}0 \\
\hline
\textbf{Rule 23} $(x^{i}_{t+1})$ & {0} & {0} & {0} & {1} & {0} & {1} & {1} & {1} \\
\textbf{Rule 77} $(x^{i}_{t+1})$& {0} & {1} & {0} & {0} & {1} & {1} & {0} & {1} \\
\textbf{Rule 178} $(x^{i}_{t+1})$& {1} & {0} & {1} & {1} & {0} & {0} & {1} & {0} \\
\textbf{Rule 232} $(x^{i}_{t+1})$& {1} & {1} & {1} & {0} & {1} & {0} & {0} & {0} \\
\hline
\end{tabular}
\end{center}
\end{table}

%\begin{table}
 %  \caption{Table captions should be placed above the tables.}\label{tab232}
 %    \begin{tabular}{lcccccccc}
 %\hline
  %    \textbf{Configuration}  & 1\textbf{1}1  & 1\textbf{1}0  & 1\textbf{0}1  & 1\textbf{0}0  & 0\textbf{1}1  & 0\textbf{1}0  & 0\textbf{0}1  & 0\textbf{0}0 \\ 
 %\hline
  %    \textbf{New State}  & \textbf{1}  & \textbf{1}  & \textbf{1}  & \textbf{0}  & \textbf{1}  & \textbf{0}  & \textbf{0}  & \textbf{0} \\ 
 %\hline
  % \end{tabular}
%\end{table}

Equations \ref{r23}--\ref{r232} serve as the fundamental rules underlying the stochastic dynamics that define the probabilistic cellular automaton presented in the next section.

\section{The Probabilistic Cellular Automata}\label{PCA}

PCA are defined by probabilistic updating rules. Consequently, for a given rule and size, the time evolution of the PCA yields different behaviours for each realization even under identical initial and boundary conditions. The updating rules of the PCA we define in this work are extensions of elementary CA rules in terms of two probabilistic parameters: $0\leq p,r \leq 1$. The first one represents the probability of keeping the current state in case of tie in the neighbors, while the second one measures the probability of adopting the state of the majority. 

\begin{table}
\begin{center}
    \caption{Probabilistic update rules: $P(x^i_{t+1}=1 \mid \text{local config.})$}\label{tabp}
    \begin{tabular}{lcccccccc}
 \hline
        \textbf{Local config.}  & 1{1}1  & 1{1}0  & 1{0}1  & 1{0}0  & 0{1}1  & 0{1}0  & 0{0}1  & 0{0}0 \\ 
 \hline
        \textbf{$P(x^i_{t+1}=1)$}  & {r}  & $p$  & r  & $1-p$  & $p$  & $1-r$  & $1-p$  & $1-r$ \\ 
 \hline
    \end{tabular}
\end{center}
\end{table}

The update of the central cell is determined probabilistically according to Bernoulli distributions with parameters $p$ and $r$. Specifically, we assume the probabilistic updating rules given in Table \ref{tabp}. %As stated above, the four extreme cases, $(p=0,r=0)$, $(p=1,r=0)$, $(p=0,r=1)$ and $(p=1,r=1)$, correspond to the rules 23, 77, 178 and 232, respectively. 
This updating probabilistic rules can also be defined using the transition probabilities for the central site to have a given state in the next step conditioned by its current state and those of its neighbors \cite{Fuks23,Fuks02}:
\[
P(x^i_{t+1} \, \mid \, x^{i-1}_t, x^i_t, x^{i+1}_t)
\]as given in Table \ref{tabp}. Note that $P(1 \, \mid  \, a, b,c) = 1 - P(1 \, \mid  \, 1-a, 1-b,1-c)$ for all $a,b,c \in \{0,1\}$. The remaining Transition Probabilities can be obtained, for $ a,b,c \in \{0,1\}$, using: $P(0 \, \mid  \, a,b,c)= 1 - P(1 \, \mid   \, a,b,c)$.

%It is straightforward to prove that probabilistic rule map, $f_{p,r}$ for $p, r \in [0,1]$, is a convex combination of the corresponding densities of rules, 23 , 77, 178 and 232 (see previous section):
%{\small
%\begin{eqnarray}\label{combi}
 %&  & f_{p,r}(x_{i-1},x_i,x_{i+1})  =   \\ \nonumber
 %&  & r \, \left[ (1-p) \, f_{178}(x_{i-1},x_i,x_{i+1}) + p \, f_{232}(x_{i-1},x_i,x_{i+1})\right] +\\ \nonumber
 %&  & (1-r) \, \left[p\, f_{77}(x_{i-1},x_i,x_{i+1})+(1-p)\, f_{23}(x_{i-1},x_i,x_{i+1})\right]
%\end{eqnarray}
%}

It is straightforward to prove that the probabilistic rule map $f_{p,r}$ for $p,r\in[0,1]$ can be expressed as:
{
\begin{eqnarray}\label{combi}
 &  & f_{p,r}(x^{i-1},x^i,x^{i+1})  =   \\ \nonumber
 &  & \mathcal{B}(r) \, \left[ (1-\mathcal{B}(p)) \, f_{178}(x^{i-1},x^i,x^{i+1}) + \mathcal{B}(p) \, f_{232}(x^{i-1},x^i,x^{i+1})\right] +\\ \nonumber
 &  & (1-\mathcal{B}(r)) \, \left[\mathcal{B}(p)\, f_{77}(x^{i-1},x^i,x^{i+1})+(1-\mathcal{B}(p))\, f_{23}(x^{i-1},x^i,x^{i+1})\right]
\end{eqnarray}
}
where $x_i \in \{0, 1\}$ and $\mathcal{B}(p)$ and $\mathcal{B}(r)$ are Bernoulli random variables with expectations $E[\mathcal{B}(p)]=p$ and $E[\mathcal{B}(r)]=r$. Consequently, the expected value of $f_{p,r}$ is the convex combination of the deterministic rules with weights $p$ and $r$.

\section{Low-size Cellular Automata: an analytical approach using Markov chains}\label{seclowsize}

The dynamics of the PCA defined in the previous section can be viewed as a Markov stochastic process over the set of all possible configurations. Let ${s_1, s_2, \ldots, s_{2^L}}$ be the set of all possible configurations of a PCA of size $L$, and let $v(t)$ be the vector whose components $v_i(t)$ represent the probability that the system is in configuration $s_i$ at time $t$. The time evolution of $v(t)$ is governed by the Stochastic Transition Matrix (STM), $A$, such that:
\begin{equation}
v(t+1)  =  v(t) \, A 
\end{equation}
where $A_{ij} = P(s_i \to s_j)$, for all $i,j=1,\ldots,2^L$, according to the probabilistic rules defined in Table \ref{tabp}. 

By recurrence, $v(t)=v(0)A^t$. Diagonalizing $A=PDP^{-1}$ gives:  
\begin{equation}
v(\infty)= v(0) \,  P \, \lim_{t\to \infty} D^t \, P^{-1} 
\end{equation}
where the limit, if it exists, can be computed directly from the eigenvalues of $A$.

% The matrix is reversible with respect to a suitable stationary distribution. Therefore, it is similar to a symmetric matrix and hence diagonalizable with real eigenvalues.

\subsection{$L=2$} 

The simplest PCA, where two cells interact has size $L=2$. In this case, consequence of the cyclic nature of the PCA, each cell is surrounded by the same cell. Therefore, for this size only majority situations appear and parameter $p$ does not affect the dynamics.

The dynamics of the PCA can be described as a Markov Chain where the states are represented as vectors, being the components of the vector $v(t) =(v_1(t),v_2(t),v_3(t),v_4(t))$ related to states as follows: $v_1  \to s_1=(0,0)$,  $v_2  \to s_2=(0,1)$, $v_3  \to s_3=(1,0)$ and $v_4  \to s_4=(1,1)$. The corresponding STM is given by: 
{
\begin{equation}A=
\left(\begin{array}{cccc}
r^{2}        & r (1-r)      & r (1-r)      & (1-r)^{2}\\
 r (1-r)      & (1-r)^{2} & r^{2}       & r (1-r)\\
 r (1-r)      & r^{2}       & (1-r)^{2} & r (1-r)\\
 (1-r)^{2} & r (1-r)     & r (1-r)       & r^{2}
\end{array}\right)
\end{equation}
} 
Note that $A$ is symmetric and then, diagonalizable. Moreover, assuming $0<r<1$, it can be shown that the STM is positive. The immediate consequence is that $A$ is irreducible, primitive and aperiodic and satisfies the strong form of the Perron-Frobenius theorem \cite{Pillaietal2005}. Therefore, $A$ has an eigenvalue 1 and the other three have modulus less than 1.  
The asymptotic vector that satisfies:
\begin{equation}
v_\infty \, A = v_\infty
\end{equation}
corresponds with the eigenvector associated to 1 and, in this particular case, it is independent of $r$:
\begin{equation}
v_\infty= \left(\frac{1}{4}, \frac{1}{4},\frac{1}{4},\frac{1}{4}\right)
\end{equation}
This global convergence from any initial condition is lost when either $r = 0$ or $r = 1$, which corresponds to a non-standard and a standard contagion, respectively.

%It is worthy to highlight that the STM is centrosymmetric which assures that the stationary distribution is symmetric.

\subsubsection{Edge $r=0$.} The entries of the STM, $A$, are: 
\begin{equation}
A_{ij} = 1 \text{ if } (i,j) \in \{(1,4),(2,2),(3,3),(4,1)\}, \text{ else } 0
\end{equation}
%\begin{equation}
%A_{ij} =
%\begin{cases}
%1, & \text{if } (i,j) \in \{(1,4),(2,2),(3,3),(4,1)\}, \\
%0, & \text{otherwise}.
%\end{cases}
%\end{equation}
whose eigenvalues are $1$ (multiplicity 3) and $-1$. Matrix $A$ is diagonalizable which allows to obtain the asymptotic vector as a function of the initial vector: $v=(v_1,v_2,v_3,v_4)$, assuming $v_1+v_2+v_3+v_4=1$:
\begin{equation}
v_{\infty}  =\left(\frac{1}{2}(v_1+v_4-(v_4-v_1)(-1)^t), v_2,v_3,\frac{1}{2}(v_1+v_4+(v_4-v_1)(-1)^t)\right)
\end{equation}
Note that this asymptotic vector exhibits oscillatory behavior whenever either $v_1$ or $v_4$ is nonzero. This remains true even when $v_1 = v_4$, meaning that both states are initially equally represented. In this case, oscillations associated with $v_1$ and $v_4$ still occur, although they are phase-shifted and cancel each other over time. For example, when the initial state is $v=(1,0,0,0)$ the final state jumps between $v_{\infty}  =(1,0,0,0)$ for $t$ even and  $v_{\infty}  =(0,0,0,1)$ for $t$ odd. This means that the PCA is periodic,  appearing a 0 row and a 1 row at each time step.

\subsubsection{Edge $r=1$.} In this case, the STM, $A$, has the following entries:

\begin{equation}
A_{ij} = 1 \text{ if } (i,j) \in \{(1,1),(2,3),(3,2),(4,4)\}, \text{ else } 0
\end{equation}
%\begin{equation}
%A_{ij} =
%\begin{cases}
%1, & \text{if } (i,j) \in \{(1,1),(2,3),(3,2),(4,4)\}, \\
%0, & \text{otherwise}.
%\end{cases}
%\end{equation}
that, as before, it is diagonalizable and has eigenvalues: 1 (multiplicity 3) and $-1$ (simple). The asymptotic vector, which also depends on the initial vector, is now:
\begin{equation}
v_{\infty} = \left( v_1, \frac{1}{2}(v_2 + v_3 + (v_2-v_3)(-1)^t), \frac{1}{2}(v_2 + v_3 - (v_2-v_3)(-1)^t), v_4\right)
\end{equation}
Contrary to the case $r=0$, a periodic situation occurs when the initial vector is either $v_3 =1$ or $v_2 =1$ (or even when $v_2=v_3$ for the same reason as before).  Note that these vectors can be viewed as equivalent under the cyclic boundary conditions.

Interestingly, this periodic situation, as well as that found for $r=0$ cannot be obtained as limit of $r \to 0$ from the general case $0<r<1$.  Similar discontinuities are going to be found for larger size PCA.

\subsection{$L=3$}

Let us solve next the lowest size for which the STM, $A$, depends on $p$ and $r$. For this case, there are $2^{3}=8$ possible configurations, which we denote as:
\[
\begin{array}{ll}
s_1 = (0,0,0),\; s_2 = (0,0,1),\; s_3 = (0,1,0),\; s_4 = (0,1,1), & \\
s_5 = (1,0,0),\; s_6 = (1,0,1),\; s_7 = (1,1,0),\; s_8 = (1,1,1). &
\end{array}
\]
For this size, the STM is:
{
\begin{equation}A=
\left(\begin{array}{cccccccc}
r^{3} & r^{2} s  & r^{2} s  & r \,s^{2} & r^{2} s  & r \,s^{2} & r \,s^{2} & s^{3}
\\
 p^{2} r  & p^{2} s  & p r q  & p s q  & p r q  & p s q  & r \,q^{2} & q^{2} s  
\\
 p^{2} r  & p r q  & p^{2} s  & p s q  & p r q  & r \,q^{2} & p s q  & q^{2} s  
\\
 q^{2} s  & p s q  & p s q  & p^{2} s  & r \,q^{2} & p r q  & p r q  & p^{2} r  
\\
 p^{2} r  & p r q  & p r q  & r \,q^{2} & p^{2} s  & p s q  & p s q  & q^{2} s  
\\
 q^{2} s  & p s q  & r \,q^{2} & p r q  & p s q  & p^{2} s  & p r q  & p^{2} r  
\\
 q^{2} s  & r \,q^{2} & p s q  & p r q  & p s q  & p r q  & p^{2} s  & p^{2} r  
\\
 s^{3} & r \,s^{2} & r \,s^{2} & r^{2} s  & r \,s^{2} & r^{2} s  & r^{2} s  & r^{3}
\end{array}\right)
\end{equation}
}
where, for the sake of simplicity, $q=1-p$ and $s=1-r$.

Again, it is not difficult to prove that, assuming $0<p,r<1$, the stochastic matrix, $A$, is diagonalizable and positive. Therefore, $A$ is irreducible, primitive and aperiodic and satisfies the strong form of the Perron-Frobenius theorem \cite{Pillaietal2005}. The main consequence is that eigenvalue 1 is simple and all other eigenvalues have modulus strictly less than 1. This is directly confirmed by the calculation of the eigenvalues, $\lambda_i$, of $A$: 
\begin{equation}\label{eigen3}
\begin{array}{l}
\lambda_1 =1 \\
\lambda_{2,3} = -3\,p^2\,r + r^3 + 3/2\,p^2 + 3\,p\,r - 3/2\,r^2 - p + r - 1/2 \pm 1/2\,\sqrt{\Delta}\\
% \lambda_3 = -3\,p^2\,r + r^3 + 3/2\,p^2 + 3\,p\,r - 3/2\,r^2 - p + r - 1/2 - 1/2\,\sqrt{\Delta} \\
\lambda_4 = -p^2 - 2\,p\,r + 3\,r^2 + 2\,p - 2\,r \\
\lambda_5 =\lambda_6 = -r + p \\
\lambda_7 =\lambda_8 = 2\,p^2 - 2\,p\,r - p + r \\
\end{array}
\end{equation}
where $\Delta = 36\,p^4\,r^2 - 24\,p^2\,r^4 + 4\,r^6 - 36\,p^4\,r - 72\,p^3\,r^2 + 48\,p^2\,r^3 + 24\,p\,r^4 - 12\,r^5 + 9\,p^4 + 60\,p^3\,r + 42\,p^2\,r^2 - 76\,p\,r^3 + r^4 - 12\,p^3 - 48\,p^2\,r + 36\,p\,r^2 + 32\,r^3 + 10\,p^2 + 4\,p\,r - 26\,r^2 - 4\,p + 4\,r + 1$. 

\begin{figure}[htbp]
  \centering
  \includegraphics[width=0.50\textwidth]{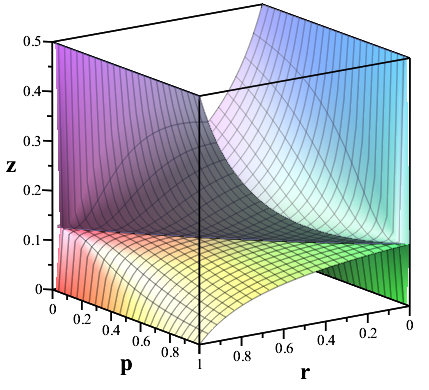}
  \caption{Stationary coordinates of the vector $v_\infty$ as functions of the parameters $p$ and $r$. Its entries can be described by the two expressions given in Eq. \ref{eq23}, where $v_1(\infty) = v_8(\infty)$ [gray] and $v_2(\infty) = v_3(\infty) = \dots = v_7(\infty)$ [black]. It is worth noting that both surfaces share a common line of tangency when $r = p$.}
  \label{estacionariogeneralL3}
\end{figure}

Except $\lambda_1$, the rest satisfy $|\lambda_i | < 1$ for all $i$ when $0<p,r<1$. This implies that the asymptotic vector, $v_\infty$, such that $v_\infty=v_\infty\,A$, is positive and unique. Its coordinates can be described by the following two expressions (see Fig. \ref{estacionariogeneralL3}):

\begin{equation}\label{eq23}
\begin{array}{lll}
v_1(\infty) &=& v_8(\infty)=\frac{p^{2}+2 p r -2 p -r +1}{2 (p^{2}+2 p r -3 r^{2}-2 p +2 r +1)}\\
v_2(\infty) &=& v_3(\infty)=\ldots=v_7(\infty)=\frac{\left(1-r\right) r}{2 \left(p^{2}+2 p r -3 r^{2}-2 p +2 r +1\right)}\\
\end{array}
\end{equation}
\normalsize

%Note that, actually, components 1 and 8 are equal, as well as components 2,3,4,5,6,7 . 

This result does not apply when either $r$ or $p$ or both take the values 0 or 1. We analyze these cases separately in the following subsections.

\subsubsection{Edge $r=0$.} This edge in the parameter space (see Fig. \ref{prsimplex}), means that, in case of majority, a non-standard contagion occurs, i.e., the central site adopts the opposite state of the majority. When a tie occurs, as before, the probability of keeping the current state of the central site is given by $p$. The STM reduces to: 
{
\begin{equation}A=
\left(\begin{array}{cccccccc}
0 & 0 & 0 & 0 & 0 & 0 & 0 & 1
\\
 0 & p^{2} & 0 & p q  & 0 & p q  & 0 & q^{2}
\\
 0 & 0 & p^{2} & p q  & 0 & 0 & p q  & q^{2}
\\
 q^{2} & p q  & p q  & p^{2} & 0 & 0 & 0 & 0
\\
 0 & 0 & 0 & 0 & p^{2} & p q  & p q  & q^{2}
\\
 q^{2} & p q  & 0 & 0 & p q  & p^{2} & 0 & 0
\\
 q^{2} & 0 & p q  & 0 & p q  & 0 & p^{2} & 0
\\
 1 & 0 & 0 & 0 & 0 & 0 & 0 & 0
\end{array}\right)
\end{equation}
}
where, as before, $q=1-p$.  

The eigenvalues of $A$, which is diagonalizable for all $p$-values, are:
\begin{equation}
\lambda= (p, p, 1, -p^2 + 2\,p, -1, 2\,p^2 - p, 2\,p^2 - p, 3\,p^2 - 2\,p)
\end{equation}
The existence of two eigenvalues with modulus 1 implies that the asymptotic vector depends on the initial vector. 
 If $0<p<1$, the PCA for $L=3$ and $r=0$, tends to the asymptotic vector:
\begin{equation}
v_{\infty}=
\left( \frac{\alpha \left(-1\right)^{t} + \beta}{2 \beta} , 0 , 0 , 0 ,  0 ,  0 ,  0 ,  \frac{-\alpha \left(-1\right)^{t} + \beta}{2 \beta} \right)
\end{equation}
where 
\begin{equation*}
\begin{array}{ll}\alpha =
& \left(-v_{7}-3 v_{8}+3 v_{1}+v_{2}+v_{3}-v_{4}+v_{5}-v_{6}\right) p^{2}+\\
&\left(2 v_{7}+2 v_{8}-2 v_{1}-2 v_{2}-2 v_{3}+2 v_{4}-2 v_{5}+2 v_{6}\right) p \\
&-v_{7}-v_{8}+v_{1}+v_{2}+v_{3}-v_{4}+v_{5}-v_{6}
\end{array}
\end{equation*}
and $\beta = 3 p^{2}- 2 p +1 $, being $v=(v_1,\ldots,v_8)$ the initial vector with $\sum_1^8 v_i=1$.

%NOTESE LAS OSCILACIONES QUE OCURREN EN ESTE CASO NO SE PUEDEN EVITAR ESCOGIENDO DETERMINADAS CONDICIONES INICIALES. SIN EMBARGO, LA INFORMACION QUE SE OBTIENE DE ESTE V INFINITO SE CORRESPONDE CON UNA MEDIA EN EL TIEMPO POR EJEMPLO SI V1 ES IGUAL A V8 IGUAL A 1/2, ENTONCES V INFINITO ES 1/2 0 0 0 0 0 0 1/2, QUE ES LA MEDIA EN EL TIEMPO DEL COMPORTAMIENTO 

This vector represents a periodic behavior for all initial conditions. For example, if $v=(0,0,0,1,0,0,0,0)$ then, the asymptotic vector simplifies:
\begin{equation}
v_{\infty}=\left(\frac{1}{2} - \frac{ (p-1)^2 (-1)^t}{2 \, (3 p^2 - 2 p + 1)},0,0,0,0,0,0, \frac{1}{2} + \frac{ (p-1)^2  (-1)^t}{2 \, (3 p^2 - 2 p + 1)}\right)
\end{equation}
which provides the probability for the PCA of being in states $s_1=(0,0,0)$ and $s_8=(1,1,1)$ at odd and even time steps. Note that,  as occurred for $L=2$, this vector cannot be obtained as a limit $r \to 0$ for the general case for any $0<p<1$. % (see Fig. \ref{r0pv41}).

%\begin{figure}[htbp]
 % \centering
 % \includegraphics[width=0.40\textwidth]{r0pv41.png}
 % \caption{Edge $r=0$. Probability of achieving the states $(0,0,0)$ and $(1,1,1)$ for any even step (red) and odd step (blue) as a function of $p$.}
 % \label{r0pv41}
%\end{figure}

\begin{figure}[htbp]
  \centering
  \includegraphics[width=0.50\textwidth]{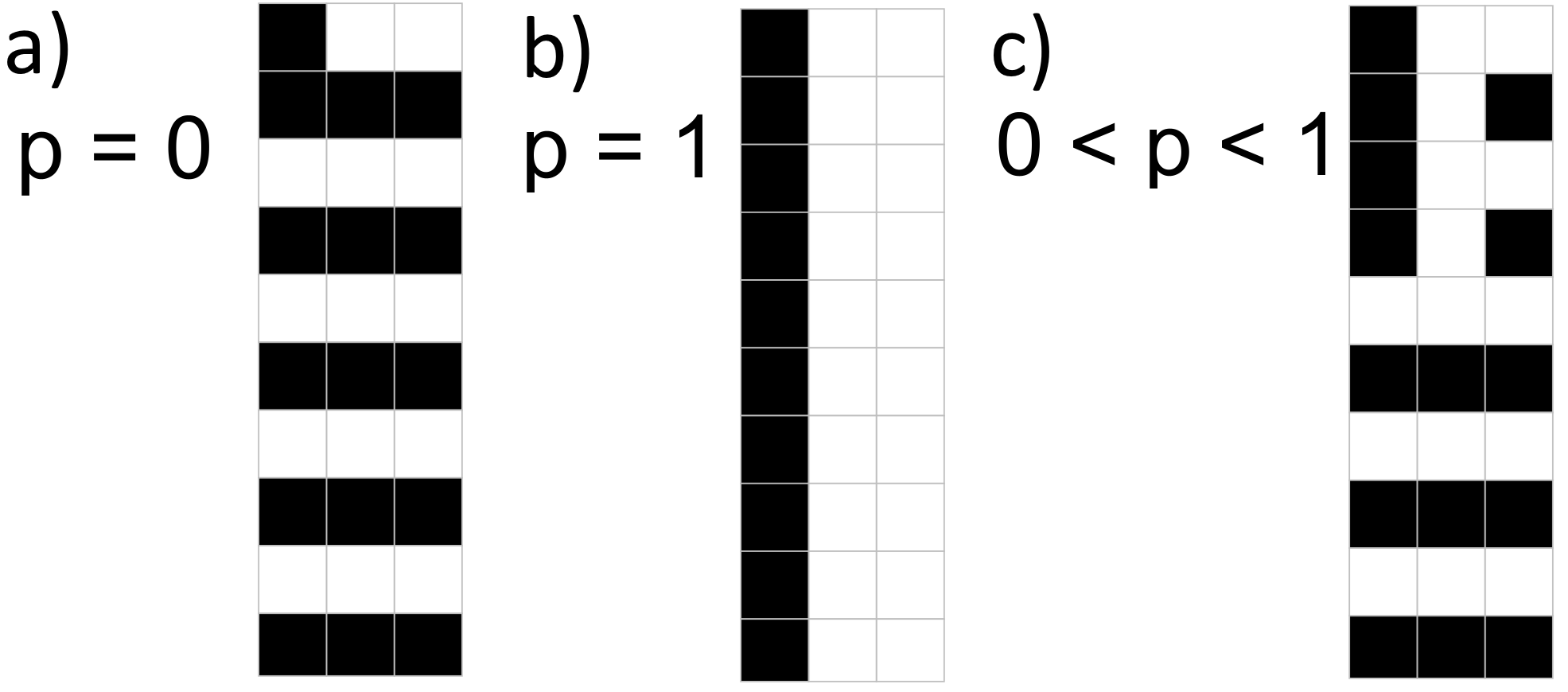}
  \caption{Examples of dynamics of the PCA in the case $r=0$, starting from $s_5=(1,0,0)$ initial configuration, with $L = 3$, white cells correspond to 0 and black cells to 1: a) $p=0$, b) $p=1$, c) $0<p<1$ (time evolves from top to bottom).}
  \label{L3ex}
\end{figure}

The limit cases $p=0$ and $p=1$ correspond to the rules 23 and 77, respectively. For the former, the asymptotic vector reduces to:
\begin{equation}v_{\infty}=
\left(\frac{1}{2}+ \frac{\alpha\left(-1\right)^{t}}{2} , 0 , 0 , 0 , 0 , 0 , 0 ,\frac{1}{2}-\frac{\alpha \left(-1\right)^{t}}{2} \right)
\end{equation}where $\alpha = 2 (v_{1}+ v_{2}+ v_{3}+ v_{5})-1$, which yields periodic behavior for all initial configurations such that $v_1,v_2,v_3$ and $v_5$ are non null (even if $v_{1}+ v_{2}+ v_{3}+ v_{5} = \frac{1}{2}$, for the same reason explained previously).  For the latter case, $p=1$, the asymptotic vector is:
\begin{equation}
v_{\infty} = \left(\frac{v_1}{2} + \frac{v_8}{2} - \frac{( v_8 - v_1)(-1)^t}{2}, v_2, v_3, v_4, v_5, v_6, v_7, \frac{v_1}{2} + \frac{v_8}{2} + \frac{( v_8 - v_1)(-1)^t}{2}\right)
\end{equation}
As before, periodic behavior can appear for particular initial conditions, specifically, all that satisfy $v_1, v_8 \neq 0$.  Again, it is worthy to remark that this vector cannot be obtained as a limit $p\to1$ from the general case with $r=0$ (see Fig. \ref{L3ex}).

%NOTESE QUE EL COMPORTAMIENTO PERIODICO OCURRE PARA TODAS LAS CONDICIONES INICIALES, NO SE PUEDE EVITAR, SIN EMBARGO. EL VECTOR ASINTOTICO SÍ RESULTA COMO UNA MEDIA EN EL TIEMPO.

\subsubsection{Edge $r=1$.} When $r=1$, in case of majority, the central site adopts the state to its neighbors (standard contagion). In case of tie, the update of the central site depends on $p$. This edge links the rules 178 ($p=0$) and 232 ($p=1$). For $0 < p < 1$, the STM is given by:
{
\begin{equation}A=
\left(\begin{array}{cccccccc}
1 & 0 & 0 & 0 & 0 & 0 & 0 & 0
\\
 p^{2} & 0 & p q  & 0 & p q  & 0 & q^{2} & 0
\\
 p^{2} & p q  & 0 & 0 & p q  & q^{2} & 0 & 0
\\
 0 & 0 & 0 & 0 & q^{2} & p q  & p q  & p^{2}
\\
 p^{2} & p q  & p q  & q^{2} & 0 & 0 & 0 & 0
\\
 0 & 0 & q^{2} & p q  & 0 & 0 & p q  & p^{2}
\\
 0 & q^{2} & 0 & p q  & 0 & p q  & 0 & p^{2}
\\
 0 & 0 & 0 & 0 & 0 & 0 & 0 & 1
\end{array}\right)
\end{equation}
}
being $q=1-p$. Its eigenvalues are:
\begin{equation}\label{eigen3r1}
\lambda= (-3 p^2 + 4 p - 1, -p^2 + 1, 1, p - 1, 2 p^2 - 3 p + 1, 1, p - 1, 2 p^2 - 3 p + 1)
\end{equation}
where, the largest eigenvalue 1 appears with double multiplicity. The rest of eigenvalues have modulus less than 1, assuming that $0<p<1$.  Then, the PCA tends to the asymptotic vector, corresponding to the subspace of eigenvalue 1: $v_\infty=(v_1(\infty), 0,0,0,0,0,0,v_8(\infty))$, with:
{
\begin{equation}\label{vec3r1}
\begin{array}{lll}
v_1(\infty) =  \frac{
(3 v_1 + 2 v_2 + 2 v_3 + v_4 + 2 v_5 + v_6 + v_7) p^2
-( 2 v_1 +  v_2 +  v_3 +  v_4 +  v_5 +  v_6 +  v_7) 2p
+ 2v_1 + v_2 + v_3 + v_4 + v_5 + v_6+ v_7 
}{3p^2 - 4p + 2}\\
v_8(\infty) =   \frac{(v_2 + v_3 + 2 v_4 + v_5 + 2 v_6 + 2 v_7 + 3 v_8) p^2
- (  v_2 +  v_3 +  v_4 +  v_5 +  v_6 +  v_7 + 2 v_8) 2p
+  v_2 + v_3 + v_4 + v_5 + v_6+ v_7 + 2v_8 
}{3p^2 - 4p + 2}
\end{array}
\end{equation}
}

As before,  $v=(v_1,\ldots,v_8)$ is the initial vector constrained to: $\sum_1^8 v_i=1$.

The asymptotic vector represents homogeneous states, either all 0 or 1, whose proportions are polynomials in $p$ with coefficients that depend on the initial vector. Contrary to the case $r=0$, no periodic behavior occurs.

Next, let us study the limit cases $p=0$ (rule 178) and $p=1$ (rule 232). For $p=0$, the transition matrix reduces to 
{
\begin{equation}
A_{ij} = 1 \text{ if } (i,j) \in \{(1,1),(2,7),(3,6),(4,5),(5,4),(6,3),(7,2),(8,8)\}, \text{ else } 0
\end{equation}
%\begin{equation}
%A_{ij} =
%\begin{cases}
%1, & \text{if } (i,j) \in \{(1,1),(2,7),(3,6),(4,5),(5,4),(6,3),(7,2),(8,8)\}, \\
%0, & \text{otherwise}.
%\end{cases}
%\end{equation}
}
whose eigenvalues are (obtained from \ref{eigen3r1} taking $p\to 0$):
\begin{equation}
\lambda= ( - 1,  1, 1, - 1,  1, 1, - 1, 1)
\end{equation}
However, the limit vector cannot be obtained from the general one, equation \ref{vec3r1}. The actual asymptotic vector, as a function of time step $t$ for any initial vector $v=(v_1,\ldots,v_8)$ such that $\sum_1^8 v_i=1$, is:
{
\begin{equation}
\begin{array}{l}
v_1(\infty) = v_{1} \\
v_2(\infty) =\left({v_{2}}+{v_{7}}-{\left(-v_{2}+v_{7}\right) \left(-1\right)^{t}}\right)/2\\ 
v_3(\infty)=\left( {v_{3}}+{v_{6}}-{\left(-v_{3}+v_{6}\right) \left(-1\right)^{t}}\right)/2\\ 
v_4(\infty)=\left( {v_{4}}+{v_{5}}-{\left(-v_{4}+v_{5}\right) \left(-1\right)^{t}}\right)/2\\ 
v_5(\infty)=\left( {v_{4}}+{v_{5}}+{\left(-v_{4}+v_{5}\right) \left(-1\right)^{t}}\right)/2\\ 
v_6(\infty)=\left( {v_{3}}+{v_{6}}+{\left(-v_{3}+v_{6}\right) \left(-1\right)^{t}}\right)/2\\ 
v_7(\infty)=\left( {v_{2}}+{v_{7}}+{\left(-v_{2}+v_{7}\right) \left(-1\right)^{t}}\right)/2\\ 
v_8(\infty) = v_{8} 
\end{array}
\end{equation} 
}
Note that this vector described oscillations for particular initial configurations,  even when $v_2=v_7$, $v_3=v_6$ and $v_4=v_5$ (see the explanation of a similar case when $r=0$ for $L=2$).  

The other limit case $p=1$, which corresponds with the vertex $(r=1,p=1)$ (rule 232), has the transition matrix: 
{
\begin{equation}
A_{ij} = 1 \text{ if } (i,j) \in \{(1,1),(2,1),(3,1),(5,1),(4,8),(6,8),(7,8),(8,8)\}, \text{ else } 0
\end{equation}
%\begin{equation}
%A_{ij} =
%\begin{cases}
%1, & \text{if } (i,j) \in \{(1,1),(2,1),(3,1),(5,1),(4,8),(6,8),(7,8),(8,8)\}, \\
%0, & \text{otherwise}.
%\end{cases}
%\end{equation}
}
with eigenvalues (again obtained from \ref{eigen3r1} taking $p\to 1$)
\begin{equation}
\lambda= ( 0 , 0, 1, 0,  0, 1, 0, 0)
\end{equation}
The asymptotic vector turns to be:
\begin{equation}
v_{\infty} = (v_1+v_2+v_3+v_5, 0,0,0,0,0,0, v_4+v_6+v_7+v_8)
\end{equation} 
which, as in previous cases with $p=1$, correspond to asymptotic homogeneous states, either $s_1=(0,0,0)$ or $s_8=(1,1,1)$, with probability distributions that depend on the initial vector $v=(v_1,\ldots,v_8)$, satisfying $\sum_{i=1}^8 \, v_i=1$.

\subsubsection{Edge $p=0$.} The other two edges of the square correspond to the values $p=0$ and $p=1$, representing,  a null probability of keeping the actual state of the central site when a tie occurs in its neighborhood and the opposite, respectively. In case of majority, the updating of the central site depends on $r$. 

For the former case, $p=0$, we can proceed as in the previous subsections calculating the STM as a function of $0<r<1$:
{
\begin{equation}A=
\left(\begin{array}{cccccccc}
r^{3} & r^{2} s  & r^{2} s  & r \,s^{2} & r^{2} s  & r \,s^{2} & r \,s^{2} & s^{3}
\\
 0 & 0 & 0 & 0 & 0 & 0 & r  & s  
\\
 0 & 0 & 0 & 0 & 0 & r  & 0 & s  
\\
 s  & 0 & 0 & 0 & r  & 0 & 0 & 0
\\
 0 & 0 & 0 & r  & 0 & 0 & 0 & s  
\\
 s  & 0 & r  & 0 & 0 & 0 & 0 & 0
\\
 s  & r  & 0 & 0 & 0 & 0 & 0 & 0
\\
 s^{3} & r \,s^{2} & r \,s^{2} & r^{2} s  & r \,s^{2} & r^{2} s  & r^{2} s  & r^{3}
\end{array}\right)
\end{equation}
}
where $s=1-r$.  We next compute its eigenvalues and corresponding eigenvectors ($A$ is still diagonalizable): 
\begin{equation}
\begin{array}{l}
\lambda_1=1 
\\
\lambda_2= 3 r^{2}-2 r  
\\
\lambda_{3,4}= r^{3}-\frac{3 r^{2}}{2}+r -\frac{1}{2}\pm\frac{\sqrt{4 r^{6}-12 r^{5}+r^{4}+32 r^{3}-26 r^{2}+4 r +1}}{2} 
%\\
%\lambda_4= r^{3}-\frac{3 r^{2}}{2}+r -\frac{1}{2}-\frac{\sqrt{4 r^{6}-12 r^{5}+r^{4}+32 r^{3}-26 r^{2}+4 r +1}}{2} 
\\
\lambda_5 = \lambda_6= r  
\\
\lambda_7 = \lambda_8= -r  
\end{array}
\end{equation}
As it can be seen, for $0<r<1$, there is a simple eigenvalue with value 1, having the rest a modulus less than 1. This allows to state that for $p=0$ and $0<r<1$, the PCA has a unique asymptotic vector, independently of the initial conditions, given by the eigenvector associated with eigenvalue 1, whose components are:
\begin{equation}
\begin{array}{l}
  v_1(\infty)   =  v_8(\infty) = \frac{1}{2 \, (3 r + 1)} \\
v_2(\infty)  = v_3(\infty) =v_4(\infty) =v_5(\infty) =v_6(\infty) =v_7(\infty) = \frac{r}{2 \, (3 r + 1)}
\end{array}
\end{equation}
%{\small
%\begin{equation}
%v_{\infty} = \left(\frac{1}{2 \, (3 r + 1)}, \frac{r}{2 \, (3 r + 1)}, \frac{r}{2 \, (3 r + 1)}, \frac{r}{2 \, (3 r + 1)}, \frac{r}{2 \, (3 r + 1)}, \frac{r}{2 \, (3 r + 1)}, \frac{r}{2 \, (3 r + 1)}, \frac{1}{2 \, (3 r + 1)}\right)
%\end{equation}
%} 
%\begin{figure}[htbp]
 % \centering
 % \includegraphics[width=0.7\textwidth]{fig5_6.png}
%  \caption{a) Probability of reaching each of the states of the PCA asymptotically for Edge $p=0$ as a function of $0<r<1$: red curve for states $(0,0,0)$ and $(1,1,1)$ and blue curve for the rest of states. b) Probability of reaching each of the states of the PCA asymptotically for $p=1$ as a function of $0<r<1$ for states $(0,0,0)$ and $(1,1,1)$ (red curve) and the rest of states (blue curve).}
 % \label{p0r}
%\end{figure}
The limit cases $r=0$ and $r=1$ have been already studied above, showing the existence of a discontinuity in these vertices. % (see Fig. \ref{p0r}a).
\subsubsection{Edge $p=1$.} The last edge to be studied for the size $L=3$ corresponds to a probability 1 of keeping the current state of the central site in case of a tie in its neighborhood. The STM in this case is (as before, $s=1-r$):
{
\begin{equation}A=
\left(\begin{array}{cccccccc}
r^{3} & r^{2} s  & r^{2} s  & r \,s^{2} & r^{2} s  & r \,s^{2} & r \,s^{2} & s^{3}
\\
 r  & s  & 0 & 0 & 0 & 0 & 0 & 0
\\
 r  & 0 & s  & 0 & 0 & 0 & 0 & 0
\\
 0 & 0 & 0 & s  & 0 & 0 & 0 & r  
\\
 r  & 0 & 0 & 0 & s  & 0 & 0 & 0
\\
 0 & 0 & 0 & 0 & 0 & s  & 0 & r  
\\
 0 & 0 & 0 & 0 & 0 & 0 & s  & r  
\\
 s^{3} & r \,s^{2} & r \,s^{2} & r^{2} s  & r \,s^{2} & r^{2} s  & r^{2} s  & r^{3}
\end{array}\right)
\end{equation}
}
Again, $A$ is diagonalizable and has the following eigenvalues: 
\begin{equation}
\begin{array}{l}
\lambda_1=1 
\\
\lambda_2 = 3 r^{2}-4 r +1 
\\
 \lambda_{3,4} = r^{3}-\frac{3 r^{2}}{2}+r \pm\frac{\sqrt{4 r^{6}-12 r^{5}+r^{4}+4 r^{3}+16 r^{2}-16 r +4}}{2} 
\\
% \lambda_4 =  r^{3}-\frac{3 r^{2}}{2}+r -\frac{\sqrt{4 r^{6}-12 r^{5}+r^{4}+4 r^{3}+16 r^{2}-16 r +4}}{2} 
%\\
 \lambda_5 =  \lambda_6 = \lambda_7 = \lambda_8 = 1-r 
\end{array}
\end{equation}
The spectrum of $A$ only contains a single eigenvalue 1 and the rest having modulus less than 1. Therefore the asymptotic vector is given by the eigenvector associated to 1, that satisfies $v_{\infty}=v_{\infty}  \, A$, and has the following components: % (see Fig. \ref{p0r}b):
\begin{equation}
\begin{array}{lll}
 v_1(\infty)   =  v_8(\infty)   =  \frac{1}{2 (4 - 3 r)} & & \\
v_2(\infty)  =  v_3(\infty) =v_4(\infty) =v_5(\infty) =v_6(\infty) =v_7(\infty) = \frac{1 - r}{2(4 - 3 r)} & &
\end{array}
\end{equation}
%{\small
%\begin{equation}
%v_{\infty} = \left(- \frac{1}{2 (3 r - 4)}, \frac{r- 1}{2(3 r - 4)}, \frac{r- 1}{2(3 r - 4)}, \frac{r- 1}{2(3 r - 4)}, \frac{r- 1}{2(3 r - 4)}, \frac{r- 1}{2(3 r - 4)}, \frac{r- 1}{2(3 r - 4)}, - \frac{1}{(2 (3 r - 4)}) \right)
%\end{equation} 
%\}
%which, as before, contains only two diferent values for the eight components 
The limit cases $r=0$ (rule 77) and $r=1$ (rule 232)  have been already studied in the previous sections.

\subsection{$L=4$}

As it will be seen, the parity in the PCA size is relevant because other stationary configurations can appear. Within this approach of exhaustive analysis of low size PCA, we present in this section the main results in comparison with the previous section.

Let us define the probabilistic vector for each time step: $v(t)=(v_1(t),v_2(t),\ldots,v_{15}(t),v_{16}(t))$, with $0 \leq v_i(t)\leq 1$ for all $t$ and $i=1,2,\ldots,16$,  that corresponds to the PCA states as follows: $v_1 \to (0,0,0,0), v_2 \to (0,0,0,1), \ldots, v_{15} \to (1,1,1,0), v_{16} \to (1,1,1,1)$, in an increasing order. 

Let us start with the strict case: $0<r,p,<1$, for which we have a similar result as for $L=3$: the Stochastic Transition Matrix of the PCA: 
{\footnotesize
\begin{equation}
\left(\begin{array}{cccccccccccccccc}
r^{4} & r^{3} s  & r^{3} s  & r^{2} s^{2} & r^{3} s  & r^{2} s^{2} & r^{2} s^{2} & r \,s^{3} & r^{3} s  & r^{2} s^{2} & r^{2} s^{2} & r \,s^{3} & r^{2} s^{2} & r \,s^{3} & r \,s^{3} & s^{4}
\\
 p^{2} r^{2} & r \,p^{2} s  & p \,r^{2} q  & p r s q  & r \,p^{2} s  & p^{2} s^{2} & p r s q  & p \,s^{2} q  & p \,r^{2} q  & p r s q  & r^{2} q^{2} & r \,q^{2} s  & p r s q  & p \,s^{2} q  & r \,q^{2} s  & s^{2} q^{2}
\\
 p^{2} r^{2} & p \,r^{2} q  & r \,p^{2} s  & p r s q  & p \,r^{2} q  & r^{2} q^{2} & p r s q  & r \,q^{2} s  & r \,p^{2} s  & p r s q  & p^{2} s^{2} & p \,s^{2} q  & p r s q  & r \,q^{2} s  & p \,s^{2} q  & s^{2} q^{2}
\\
 p^{2} q^{2} & p^{3} q  & p^{3} q  & p^{4} & p \,q^{3} & p^{2} q^{2} & p^{2} q^{2} & p^{3} q  & p \,q^{3} & p^{2} q^{2} & p^{2} q^{2} & p^{3} q  & q^{4} & p \,q^{3} & p \,q^{3} & p^{2} q^{2}
\\
 p^{2} r^{2} & r \,p^{2} s  & p \,r^{2} q  & p r s q  & r \,p^{2} s  & p^{2} s^{2} & p r s q  & p \,s^{2} q  & p \,r^{2} q  & p r s q  & r^{2} q^{2} & r \,q^{2} s  & p r s q  & p \,s^{2} q  & r \,q^{2} s  & s^{2} q^{2}
\\
 r^{2} s^{2} & r \,s^{3} & r^{3} s  & r^{2} s^{2} & r \,s^{3} & s^{4} & r^{2} s^{2} & r \,s^{3} & r^{3} s  & r^{2} s^{2} & r^{4} & r^{3} s  & r^{2} s^{2} & r \,s^{3} & r^{3} s  & r^{2} s^{2}
\\
 p^{2} q^{2} & p \,q^{3} & p^{3} q  & p^{2} q^{2} & p^{3} q  & p^{2} q^{2} & p^{4} & p^{3} q  & p \,q^{3} & q^{4} & p^{2} q^{2} & p \,q^{3} & p^{2} q^{2} & p \,q^{3} & p^{3} q  & p^{2} q^{2}
\\
 s^{2} q^{2} & p \,s^{2} q  & r \,q^{2} s  & p r s q  & p \,s^{2} q  & p^{2} s^{2} & p r s q  & r \,p^{2} s  & r \,q^{2} s  & p r s q  & r^{2} q^{2} & p \,r^{2} q  & p r s q  & r \,p^{2} s  & p \,r^{2} q  & p^{2} r^{2}
\\
 p^{2} r^{2} & p \,r^{2} q  & r \,p^{2} s  & p r s q  & p \,r^{2} q  & r^{2} q^{2} & p r s q  & r \,q^{2} s  & r \,p^{2} s  & p r s q  & p^{2} s^{2} & p \,s^{2} q  & p r s q  & r \,q^{2} s  & p \,s^{2} q  & s^{2} q^{2}
\\
 p^{2} q^{2} & p^{3} q  & p \,q^{3} & p^{2} q^{2} & p \,q^{3} & p^{2} q^{2} & q^{4} & p \,q^{3} & p^{3} q  & p^{4} & p^{2} q^{2} & p^{3} q  & p^{2} q^{2} & p^{3} q  & p \,q^{3} & p^{2} q^{2}
\\
 r^{2} s^{2} & r^{3} s  & r \,s^{3} & r^{2} s^{2} & r^{3} s  & r^{4} & r^{2} s^{2} & r^{3} s  & r \,s^{3} & r^{2} s^{2} & s^{4} & r \,s^{3} & r^{2} s^{2} & r^{3} s  & r \,s^{3} & r^{2} s^{2}
\\
 s^{2} q^{2} & r \,q^{2} s  & p \,s^{2} q  & p r s q  & r \,q^{2} s  & r^{2} q^{2} & p r s q  & p \,r^{2} q  & p \,s^{2} q  & p r s q  & p^{2} s^{2} & r \,p^{2} s  & p r s q  & p \,r^{2} q  & r \,p^{2} s  & p^{2} r^{2}
\\
 p^{2} q^{2} & p \,q^{3} & p \,q^{3} & q^{4} & p^{3} q  & p^{2} q^{2} & p^{2} q^{2} & p \,q^{3} & p^{3} q  & p^{2} q^{2} & p^{2} q^{2} & p \,q^{3} & p^{4} & p^{3} q  & p^{3} q  & p^{2} q^{2}
\\
 s^{2} q^{2} & p \,s^{2} q  & r \,q^{2} s  & p r s q  & p \,s^{2} q  & p^{2} s^{2} & p r s q  & r \,p^{2} s  & r \,q^{2} s  & p r s q  & r^{2} q^{2} & p \,r^{2} q  & p r s q  & r \,p^{2} s  & p \,r^{2} q  & p^{2} r^{2}
\\
 s^{2} q^{2} & r \,q^{2} s  & p \,s^{2} q  & p r s q  & r \,q^{2} s  & r^{2} q^{2} & p r s q  & p \,r^{2} q  & p \,s^{2} q  & p r s q  & p^{2} s^{2} & r \,p^{2} s  & p r s q  & p \,r^{2} q  & r \,p^{2} s  & p^{2} r^{2}
\\
 s^{4} & r \,s^{3} & r \,s^{3} & r^{2} s^{2} & r \,s^{3} & r^{2} s^{2} & r^{2} s^{2} & r^{3} s  & r \,s^{3} & r^{2} s^{2} & r^{2} s^{2} & r^{3} s  & r^{2} s^{2} & r^{3} s  & r^{3} s  & r^{4}
\end{array}\right)
\end{equation}
}
where, as before, $s=1-r$ and $q=1-p$, is positive. The proof is straigtforward: as this matrix has  $2^4 \times 2^4$ entries, it can be analyzed one by one which enable to show that all are positive.  As stated above, the main consequence is that $A$ is irreducible and primitive and satisfy the strong form of the Perron-Frobenius theorem which assures the existence of an unique eigenvalue 1 with multiplicity 1. Moreover, the rest of eigenvalues as modulus less than 1. 

The asymptotic behavior of the Markov process is then, given by the eigenvector $v(\infty)$, that satisfies $v(\infty) \, A = v(\infty)$ for all initial conditions. 

{\small
\begin{eqnarray}
v_1(\infty) &=& -\frac{q \gamma p}{4 \alpha}\\
v_2(\infty) &=& -\frac{p r s q \left(p^{2}+r^{2}-p -r +1\right)}{2 \alpha}\\
v_3(\infty) &=& -\frac{p r s q \left(p^{2}+r^{2}-p -r +1\right)}{2 \alpha}\\
v_4(\infty) &=& \frac{s^{2} \left(2 p^{2}-2 r^{2}-2 p +2 r -1\right) r^{2}}{\beta}\\
v_5(\infty) &=& -\frac{p r s q \left(p^{2}+r^{2}-p -r +1\right)}{2 \alpha}\\
v_6(\infty) &=& -\frac{q \delta p}{4 \alpha}\\
v_7(\infty) &=& \frac{s^{2} \left(2 p^{2}-2 r^{2}-2 p +2 r -1\right) r^{2}}{\beta}\\
v_8(\infty) &=& -\frac{p r s q \left(p^{2}+r^{2}-p -r +1\right)}{2 \alpha}\\
v_9(\infty) &=& -\frac{p r s q \left(p^{2}+r^{2}-p -r +1\right)}{2 \alpha}\\
v_10(\infty) &=& \frac{s^{2} \left(2 p^{2}-2 r^{2}-2 p +2 r -1\right) r^{2}}{\beta}\\
v_11(\infty) &=& -\frac{q \delta p}{4 \alpha}\\
v_12(\infty) &=& -\frac{p r s q \left(p^{2}+r^{2}-p -r +1\right)}{2 \alpha}\\
v_13(\infty) &=& \frac{s^{2} \left(2 p^{2}-2 r^{2}-2 p +2 r -1\right) r^{2}}{\beta}\\
v_14(\infty) &=& -\frac{p r s q \left(p^{2}+r^{2}-p -r +1\right)}{2 \alpha}\\
v_15(\infty) &=& -\frac{p r s q \left(p^{2}+r^{2}-p -r +1\right)}{2 \alpha}\\
v_16(\infty) &=& -\frac{q \gamma p}{4 \alpha}
\end{eqnarray}}
where 

\begin{figure}[htbp]
  \centering
  \includegraphics[width=0.60\textwidth]{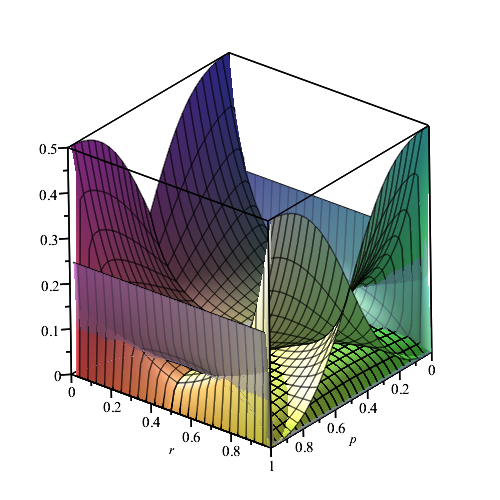}
  \caption{Probabilistic components of tha asymptotic vector as a function of $0<r,p<1$ for $L=4$.}
  \label{L4vinfty}
\end{figure}

{

$\gamma = 2 p^{4}+4 p^{3} r +6 p^{2} r^{2}+4 r^{3} p -6 p^{3}-12 p^{2} r -12 r^{2} p -2 r^{3}+8 p^{2}+14 p r +5 r^{2}-6 p -5 r +2$,

$\delta = 2 p^{4}-4 p^{3} r +6 p^{2} r^{2}-4 r^{3} p -2 p^{3}+2 r^{3}+2 p^{2}-2 p r -r^{2}+r $,

$\alpha = 2 p^{6}+2 p^{4} r^{2}-2 p^{2} r^{4}-2 r^{6}-6 p^{5}-2 p^{4} r -4 p^{3} r^{2}+4 p^{2} r^{3}+2 r^{4} p +6 r^{5}+9 p^{4}+4 p^{3} r -2 p^{2} r^{2}-4 r^{3} p -7 r^{4}-8 p^{3}+4 r^{2} p +4 r^{3}+4 p^{2}-2 p r -r^{2}-p $,

$\beta = 8 p^{6}+8 p^{4} r^{2}-8 p^{2} r^{4}-8 r^{6}-24 p^{5}-8 p^{4} r -16 p^{3} r^{2}+16 p^{2} r^{3}+8 r^{4} p +24 r^{5}+36 p^{4}+16 p^{3} r -8 p^{2} r^{2}-16 r^{3} p -28 r^{4}-32 p^{3}+16 r^{2} p +16 r^{3}+16 p^{2}-8 p r -4 r^{2}-4 p$

}
Note that, due to symmetry, there are components that coincide. In particular,  $v_1(\infty) = v_{16}(\infty)$ and $10, 4,  7, 13$ and $ 14, 15, 2 ,3 ,5 ,8 , 9$ and $ 6 $.
Figure \ref{L4vinfty} depicts the graphs of all these components as a function of $r$ and $p$. 

The positivity condition that assures the existence of a unique asymptotic probabilistic vector is lost if either $r$ or $p$ or both take the values 0 and 1. Geometrically, these cases means to reside in the edges or vertices of the square $0 \leq r,p, \leq 1$, as depicted in Fig.\ref{prsimplex}. In the following, we consider these situations sequentially and highlight the differences with respect the general case.

\subsection{Edge $r = 0$}

When $r=0$ the Stochastic Transition Matrix simplifies to: 
{\scriptsize
\begin{equation}
A  = \left(\begin{array}{cccccccccccccccc}
0 & 0 & 0 & 0 & 0 & 0 & 0 & 0 & 0 & 0 & 0 & 0 & 0 & 0 & 0 & 1 
\\
 0 & 0 & 0 & 0 & 0 & p^{2} & 0 & p q  & 0 & 0 & 0 & 0 & 0 & p q  & 0 & q^{2} 
\\
 0 & 0 & 0 & 0 & 0 & 0 & 0 & 0 & 0 & 0 & p^{2} & p q  & 0 & 0 & p q  & q^{2} 
\\
 p^{2} q^{2} & p^{3} q  & p^{3} q  & p^{4} & p \,q^{3} & p^{2} q^{2} & p^{2} q^{2} & p^{3} q  & p \,q^{3} & p^{2} q^{2} & p^{2} q^{2} & p^{3} q  & q^{4} & p \,q^{3} & p \,q^{3} & p^{2} q^{2} 
\\
 0 & 0 & 0 & 0 & 0 & p^{2} & 0 & p q  & 0 & 0 & 0 & 0 & 0 & p q  & 0 & q^{2} 
\\
 0 & 0 & 0 & 0 & 0 & 1 & 0 & 0 & 0 & 0 & 0 & 0 & 0 & 0 & 0 & 0 
\\
 p^{2} q^{2} & p \,q^{3} & p^{3} q  & p^{2} q^{2} & p^{3} q  & p^{2} q^{2} & p^{4} & p^{3} q  & p \,q^{3} & q^{4} & p^{2} q^{2} & p \,q^{3} & p^{2} q^{2} & p \,q^{3} & p^{3} q  & p^{2} q^{2} 
\\
 q^{2} & p q  & 0 & 0 & p q  & p^{2} & 0 & 0 & 0 & 0 & 0 & 0 & 0 & 0 & 0 & 0 
\\
 0 & 0 & 0 & 0 & 0 & 0 & 0 & 0 & 0 & 0 & p^{2} & p q  & 0 & 0 & p q  & q^{2} 
\\
 p^{2} q^{2} & p^{3} q  & p \,q^{3} & p^{2} q^{2} & p \,q^{3} & p^{2} q^{2} & q^{4} & p \,q^{3} & p^{3} q  & p^{4} & p^{2} q^{2} & p^{3} q  & p^{2} q^{2} & p^{3} q  & p \,q^{3} & p^{2} q^{2} 
\\
 0 & 0 & 0 & 0 & 0 & 0 & 0 & 0 & 0 & 0 & 1 & 0 & 0 & 0 & 0 & 0 
\\
 q^{2} & 0 & p q  & 0 & 0 & 0 & 0 & 0 & p q  & 0 & p^{2} & 0 & 0 & 0 & 0 & 0 
\\
 p^{2} q^{2} & p \,q^{3} & p \,q^{3} & q^{4} & p^{3} q  & p^{2} q^{2} & p^{2} q^{2} & p \,q^{3} & p^{3} q  & p^{2} q^{2} & p^{2} q^{2} & p \,q^{3} & p^{4} & p^{3} q  & p^{3} q  & p^{2} q^{2} 
\\
 q^{2} & p q  & 0 & 0 & p q  & p^{2} & 0 & 0 & 0 & 0 & 0 & 0 & 0 & 0 & 0 & 0 
\\
 q^{2} & 0 & p q  & 0 & 0 & 0 & 0 & 0 & p q  & 0 & p^{2} & 0 & 0 & 0 & 0 & 0 
\\
 1 & 0 & 0 & 0 & 0 & 0 & 0 & 0 & 0 & 0 & 0 & 0 & 0 & 0 & 0 & 0 
\end{array}\right)
\end{equation}
}
whose eigenvalues are, in general, functions of parameter $p$. Concretely, 
$\lambda_1= 0$ with multiplicity 3, $\lambda_2= 1$ with multiplicity 3, $\lambda_3= -1$ simple, $\lambda_4 = \,p^3-6 p\,^2+4\,p-1$, with multiplicity 2, $\lambda_5=-2\,p^2+2\,p$, double, $\lambda_6=2\,p^2- 2\,p$, double and $\lambda_7=4\,p^2-4\,p+1$, simple. 
Note that all the eigenvalues with $p$-dependence have modulus less than one, which means that they tends to 0 in the asymptotic limit:
\begin{equation}
v(\infty) = v \, P \lim_{n\to\infty}D^n P^{-1}
\end{equation}
being $v$ the initial vector, $P$ the matrix of basis change and $D$ the matrix whose diagonal contains the eigenvalues of $A$.  Note that, strictily speaking, this limit does not exist since the eigenvalues $-1$ yields two alternating vector, for odd and even $n$-values. The asymptotic vector components are:
{\tiny
$
v_1(\infty)  = \frac{1}{8 (p^{2}-p +1) (p^{2}-p +\frac{1}{2})}(-2 (p^{2}-p +1) ((v_{8}+v_{12}+v_{14}+v_{15}+2 v_{16}-2 v_{1}) p^{2}+(-2 v_{8}-2 v_{12}-2 v_{14}-2 v_{15}-2 v_{16}+2 v_{1}) p +(-v_{9}-v_{2}-v_{3}-v_{5}) q^{2}+v_{8}+v_{12}+v_{14}+v_{15}+v_{16}-v_{1}) (-1)^{n} +(2 v_{7}+2 v_{8}+2 v_{9}+2 v_{10}+2 v_{12}+2 v_{13}+2 v_{14}+2 v_{15}+4 v_{16}+4 v_{1}+2 v_{2}+2 v_{3}+2 v_{4}+2 v_{5}) p^{4}+(-8 v_{7}-6 v_{8}-6 v_{9}-8 v_{10}-6 v_{12}-8 v_{13}-6 v_{14}-6 v_{15}-8 v_{16}-8 v_{1}-6 v_{2}-6 v_{3}-8 v_{4}-6 v_{5}) p^{3} +(11 v_{7}+8 v_{8}+8 v_{9}+11 v_{10}+8 v_{12}+11 v_{13}+8 v_{14}+8 v_{15}+10 v_{16}+10 v_{1}+8 v_{2}+8 v_{3}+11 v_{4}+8 v_{5}) p^{2}+(-7 v_{7}-6 v_{8}-6 v_{9}-7 v_{10}-6 v_{12}-7 v_{13}-6 v_{14}-6 v_{15}-6 v_{16}-6 v_{1}-6 v_{2}-6 v_{3}-7 v_{4}-6 v_{5}) p +2 v_{7}+2 v_{8}+2 v_{9}+2 v_{10}+2 v_{12}+2 v_{13}+2 v_{14}+2 v_{15}+2 v_{16}+2 v_{1}+2 v_{2}+2 v_{3}+2 v_{4}+2 v_{5})
$

$v_6(\infty)=\frac{1}{8 (p^{2}-p +\frac{1}{2}) (p^{2}-p +1)}(2 (2 v_{14}+2 v_{2}+v_{4}+2 v_{5}+4 v_{6}+v_{7}+2 v_{8}+v_{10}+v_{13}) p^{4}+4 (-v_{14}-v_{2}-v_{5}-4 v_{6}-v_{8}) p^{3}+(4 v_{14}+4 v_{2}-v_{4}+4 v_{5}+20 v_{6}-v_{7}+4 v_{8}-v_{10}-v_{13}) p^{2}+(v_{4}-12 v_{6}+v_{7}+v_{10}+v_{13}) p +4 v_{6})$,

$v_{11}(\infty)=\frac{1}{8 (p^{2}-p +\frac{1}{2}) (p^{2}-p +1)} (2 (2 v_{15}+2 v_{3}+v_{4}+v_{7}+2 v_{9}+v_{10}+4 v_{11}+2 v_{12}+v_{13}) p^{4}+4 (-v_{15}-v_{3}-v_{9}-4 v_{11}-v_{12}) p^{3}+(4 v_{15}+4 v_{3}-v_{4}-v_{7}+4 v_{9}-v_{10}+20 v_{11}+4 v_{12}-v_{13}) p^{2}+(v_{4}+v_{7}+v_{10}-12 v_{11}+v_{13}) p +4 v_{11}) $,

$v_{16}(\infty)=\frac{1}{8 (p^{2}-p +1) (p^{2}-p +\frac{1}{2})}(2 (p^{2}-p +1) ((v_{8}+v_{12}+v_{14}+v_{15}+2 v_{16}-2 v_{1}) p^{2}+(-2 v_{8}-2 v_{12}-2 v_{14}-2 v_{15}-2 v_{16}+2 v_{1}) p +(-v_{9}-v_{2}-v_{3}-v_{5}) q^{2}+v_{8}+v_{12}+v_{14}+v_{15}+v_{16}-v_{1}) (-1)^{n}+(2 v_{7}+2 v_{8}+2 v_{9}+2 v_{10}+2 v_{12}+2 v_{13}+2 v_{14}+2 v_{15}+4 v_{16}+4 v_{1}+2 v_{2}+2 v_{3}+2 v_{4}+2 v_{5}) p^{4}+(-8 v_{7}-6 v_{8}-6 v_{9}-8 v_{10}-6 v_{12}-8 v_{13}-6 v_{14}-6 v_{15}-8 v_{16}-8 v_{1}-6 v_{2}-6 v_{3}-8 v_{4}-6 v_{5}) p^{3}+(11 v_{7}+8 v_{8}+8 v_{9}+11 v_{10}+8 v_{12}+11 v_{13}+8 v_{14}+8 v_{15}+10 v_{16}+10 v_{1}+8 v_{2}+8 v_{3}+11 v_{4}+8 v_{5}) p^{2}+(-7 v_{7}-6 v_{8}-6 v_{9}-7 v_{10}-6 v_{12}-7 v_{13}-6 v_{14}-6 v_{15}-6 v_{16}-6 v_{1}-6 v_{2}-6 v_{3}-7 v_{4}-6 v_{5}) p +2 v_{7}+2 v_{8}+2 v_{9}+2 v_{10}+2 v_{12}+2 v_{13}+2 v_{14}+2 v_{15}+2 v_{16}+2 v_{1}+2 v_{2}+2 v_{3}+2 v_{4}+2 v_{5})$
}
being the rest of coordinates of $v(\infty)$ strictly null.

Note that the time step $n$ only appears in coordinates $v_1(\infty)$ and $v_{16}(\infty)$, which means that, in this case, periodicity exclusively affects to states $(0,0,0,0)$ and $(1,1,1,1)$ and that, for certain initial configurations, this periodic behavior does not appear. For example, this is the case when:
{\small
\begin{equation}
\begin{aligned}
&(v_{8}+v_{12}+v_{14}+v_{15}+2 v_{16}-2 v_{1}) p^{2} \\
&+(-2 v_{8}-2 v_{12}-2 v_{14}-2 v_{15}-2 v_{16}+2 v_{1}) p \\
&+(-v_{9}-v_{2}-v_{3}-v_{5}) q^{2} \\
&+v_{8}+v_{12}+v_{14}+v_{15}+v_{16}-v_{1} = 0
\end{aligned}
\end{equation}
}
On the contrary, it is easy to find conditions for periodicity, for instance: $v_1=1$ and $v_i=0$ for $i=2,3,\ldots,16$, independently of the $p$-value.

Another interesting behaviour appear when $v_4=1$, being the rest of coordinates null. In this case, the asymptotic behaviour depends on $p$. For $p=0$ and $p=1$ alternation appears between $v_1$ and $v_{16}$ but, on the contrary, for $0<p<1$ asymptotic vector is:
$v_1(\infty)=v_{16}(\infty)=\frac{p^2 - 3p + 2}{4 p^2 - 4 p + 4}$ and $v_6(\infty)=v_{11}(\infty)= \frac{(p + 1)\,p}{4p^2 - 4\,p + 4}$. 

It is also interesting to compute the limits $p\to 0$ and $p\to 1$ of the asymptotic vector in order to compare with the deterministic rules studied in the next subsections.  The former limit yields:
{\tiny
\begin{equation}
\begin{aligned}
v_1(\infty)
&=
\frac{
\left(
v_{9}-v_{12}-v_{14}-v_{15}-v_{16}
+v_{1}+v_{2}+v_{3}+v_{5}-v_{8}
\right)(-1)^n
}{2}
\\
&\quad
+\frac{
v_{9}+v_{10}+v_{12}+v_{13}+v_{14}+v_{15}+v_{16}
+v_{1}+v_{2}+v_{3}+v_{4}+v_{5}+v_{7}+v_{8}
}{2},
\\[2mm]
v_6(\infty)
&= v_6,
\\[1mm]
v_{11}(\infty)
&= v_{11},
\\[2mm]
v_{16}(\infty)
&=
(-1)^n
\left(
-\frac{v_{2}}{2}-\frac{v_{3}}{2}-\frac{v_{5}}{2}
-\frac{v_{9}}{2}-\frac{v_{1}}{2}
+\frac{v_{8}}{2}+\frac{v_{12}}{2}
+\frac{v_{14}}{2}+\frac{v_{15}}{2}
+\frac{v_{16}}{2}
\right)
\\
&\quad
+\frac{
v_{4}+v_{13}+v_{10}+v_{7}
+v_{9}+v_{2}+v_{3}+v_{5}
+v_{14}+v_{15}+v_{8}
+v_{12}+v_{16}+v_{1}
}{2}.
\end{aligned}
\end{equation}
}
The second limit at $p=1$ turns out:
{\small
\begin{equation}
\begin{array}{lll}
v_1(\infty) & = & \frac{\left(v_{1}-v_{16}\right) \left(-1\right)^{n}}{2}+\frac{v_{1}}{2}+\frac{v_{16}}{2} \\
v_6(\infty) & = & v_{2}+\frac{v_{4}}{2}+v_{5}+v_{6}+\frac{v_{7}}{2}+v_{8}+\frac{v_{10}}{2}+\frac{v_{13}}{2}+v_{14} \\
v_{11}(\infty) & = & v_{3}+\frac{v_{4}}{2}+\frac{v_{7}}{2}+v_{9}+\frac{v_{10}}{2}+v_{11}+v_{12}+\frac{v_{13}}{2}+v_{15} \\
v_{16}(\infty) & = & -\frac{\left(-1\right)^{n} v_{1}}{2}+\frac{\left(-1\right)^{n} v_{16}}{2}+\frac{v_{16}}{2}+\frac{v_{1}}{2}
\end{array}
\end{equation}
}

{\bf Vertex $p=0$}.  As occurs for size $L=3$, the vertices of the square are limit cases for the computation of the aymptotic behavior. In this subsection, we first studied the vertex $(r=0,p=0)$ that corresponds with the deterministic rule 23.  As we will see, the most relevant result concerns with the discontinuity in the asymptotic vector that appears as $p$ tends to 0 due to the appearance of multiple eigenvalues with modulus 1 for the Stochastic Transition Matrix:
{\small
\begin{equation}
A  = \left(\begin{array}{cccccccccccccccc}
0 & 0 & 0 & 0 & 0 & 0 & 0 & 0 & 0 & 0 & 0 & 0 & 0 & 0 & 0 & 1 
\\
 0 & 0 & 0 & 0 & 0 & 0 & 0 & 0 & 0 & 0 & 0 & 0 & 0 & 0 & 0 & 1 
\\
 0 & 0 & 0 & 0 & 0 & 0 & 0 & 0 & 0 & 0 & 0 & 0 & 0 & 0 & 0 & 1 
\\
 0 & 0 & 0 & 0 & 0 & 0 & 0 & 0 & 0 & 0 & 0 & 0 & 1 & 0 & 0 & 0 
\\
 0 & 0 & 0 & 0 & 0 & 0 & 0 & 0 & 0 & 0 & 0 & 0 & 0 & 0 & 0 & 1 
\\
 0 & 0 & 0 & 0 & 0 & 1 & 0 & 0 & 0 & 0 & 0 & 0 & 0 & 0 & 0 & 0 
\\
 0 & 0 & 0 & 0 & 0 & 0 & 0 & 0 & 0 & 1 & 0 & 0 & 0 & 0 & 0 & 0 
\\
 1 & 0 & 0 & 0 & 0 & 0 & 0 & 0 & 0 & 0 & 0 & 0 & 0 & 0 & 0 & 0 
\\
 0 & 0 & 0 & 0 & 0 & 0 & 0 & 0 & 0 & 0 & 0 & 0 & 0 & 0 & 0 & 1 
\\
 0 & 0 & 0 & 0 & 0 & 0 & 1 & 0 & 0 & 0 & 0 & 0 & 0 & 0 & 0 & 0 
\\
 0 & 0 & 0 & 0 & 0 & 0 & 0 & 0 & 0 & 0 & 1 & 0 & 0 & 0 & 0 & 0 
\\
 1 & 0 & 0 & 0 & 0 & 0 & 0 & 0 & 0 & 0 & 0 & 0 & 0 & 0 & 0 & 0 
\\
 0 & 0 & 0 & 1 & 0 & 0 & 0 & 0 & 0 & 0 & 0 & 0 & 0 & 0 & 0 & 0 
\\
 1 & 0 & 0 & 0 & 0 & 0 & 0 & 0 & 0 & 0 & 0 & 0 & 0 & 0 & 0 & 0 
\\
 1 & 0 & 0 & 0 & 0 & 0 & 0 & 0 & 0 & 0 & 0 & 0 & 0 & 0 & 0 & 0 
\\
 1 & 0 & 0 & 0 & 0 & 0 & 0 & 0 & 0 & 0 & 0 & 0 & 0 & 0 & 0 & 0 
\end{array}\right)
\end{equation}
}
Specifically, the eigenvalues of $A$ are: 1 with multiplicity 5, -1 with multiplicity 3 and 0 with multiplicity 8. The main consequence of this high multiplicity is the existence of multistability situations where the asymptotic state depends on the initial conditions.  

If $v=(v_1,v_2, \ldots,v_{15},v_{16})$ is the initial vector then, applying the diagonalization of $A$, the asymptotic vector turns out to be (note that, strictly speaking, this is a multistate limit due to the possible alternation between states $(0,0,0,0)$ and $(1,1,1,1)$):
{\scriptsize
\begin{equation}
\left(\begin{array}{c}
-\frac{\left(-v_{1}-v_{2}-v_{3}-v_{5}+v_{8}-v_{9}+v_{12}+v_{14}+v_{15}+v_{16}\right) \left(-1\right)^{n}}{2}+\frac{v_{1}}{2}+\frac{v_{2}}{2}
\\
+\frac{v_{3}}{2}+\frac{v_{5}}{2}+\frac{v_{8}}{2}+\frac{v_{9}}{2}+\frac{v_{12}}{2}+\frac{v_{14}}{2}+\frac{v_{15}}{2}+\frac{v_{16}}{2} \\  0 \\ 0 \\ -\frac{\left(-v_{4}+v_{13}\right) \left(-1\right)^{n}}{2}+\frac{v_{4}}{2}+\frac{v_{13}}{2} \\ 0 \\ v_{6} \\  -\frac{\left(-v_{7}+v_{10}\right) \left(-1\right)^{n}}{2}+\frac{v_{7}}{2}+\frac{v_{10}}{2} \\ 0 \\ 0 \\ \frac{\left(-v_{7}+v_{10}\right) \left(-1\right)^{n}}{2}+\frac{v_{7}}{2}+\frac{v_{10}}{2} \\ v_{11} \\ 0 \\ \frac{\left(-v_{4}+v_{13}\right) \left(-1\right)^{n}}{2}+\frac{v_{4}}{2}+\frac{v_{13}}{2} \\ 0 \\ 0 \\ \frac{\left(-v_{1}-v_{2}-v_{3}-v_{5}+v_{8}-v_{9}+v_{12}+v_{14}+v_{15}+v_{16}\right) \left(-1\right)^{n}}{2}+\frac{v_{1}}{2}+\frac{v_{2}}{2}+
\\
\frac{v_{3}}{2}+\frac{v_{5}}{2}+\frac{v_{8}}{2}+\frac{v_{9}}{2}+\frac{v_{12}}{2}+\frac{v_{14}}{2}+\frac{v_{15}}{2}+\frac{v_{16}}{2} 
\end{array}\right)
\end{equation}
}

{\bf Vertex $p=1$}.  As in the previous case, the dynamics of the PCA at this vertex ($p\to 1$) is discontinuous with respect to the case $0 < p < 1$. IIt is easy to see from the Stochastic Transition Matrix is:
{\small
\begin{equation}
A  = \left(\begin{array}{cccccccccccccccc}
0 & 0 & 0 & 0 & 0 & 0 & 0 & 0 & 0 & 0 & 0 & 0 & 0 & 0 & 0 & 1 
\\
 0 & 0 & 0 & 0 & 0 & 1 & 0 & 0 & 0 & 0 & 0 & 0 & 0 & 0 & 0 & 0 
\\
 0 & 0 & 0 & 0 & 0 & 0 & 0 & 0 & 0 & 0 & 1 & 0 & 0 & 0 & 0 & 0 
\\
 0 & 0 & 0 & 1 & 0 & 0 & 0 & 0 & 0 & 0 & 0 & 0 & 0 & 0 & 0 & 0 
\\
 0 & 0 & 0 & 0 & 0 & 1 & 0 & 0 & 0 & 0 & 0 & 0 & 0 & 0 & 0 & 0 
\\
 0 & 0 & 0 & 0 & 0 & 1 & 0 & 0 & 0 & 0 & 0 & 0 & 0 & 0 & 0 & 0 
\\
 0 & 0 & 0 & 0 & 0 & 0 & 1 & 0 & 0 & 0 & 0 & 0 & 0 & 0 & 0 & 0 
\\
 0 & 0 & 0 & 0 & 0 & 1 & 0 & 0 & 0 & 0 & 0 & 0 & 0 & 0 & 0 & 0 
\\
 0 & 0 & 0 & 0 & 0 & 0 & 0 & 0 & 0 & 0 & 1 & 0 & 0 & 0 & 0 & 0 
\\
 0 & 0 & 0 & 0 & 0 & 0 & 0 & 0 & 0 & 1 & 0 & 0 & 0 & 0 & 0 & 0 
\\
 0 & 0 & 0 & 0 & 0 & 0 & 0 & 0 & 0 & 0 & 1 & 0 & 0 & 0 & 0 & 0 
\\
 0 & 0 & 0 & 0 & 0 & 0 & 0 & 0 & 0 & 0 & 1 & 0 & 0 & 0 & 0 & 0 
\\
 0 & 0 & 0 & 0 & 0 & 0 & 0 & 0 & 0 & 0 & 0 & 0 & 1 & 0 & 0 & 0 
\\
 0 & 0 & 0 & 0 & 0 & 1 & 0 & 0 & 0 & 0 & 0 & 0 & 0 & 0 & 0 & 0 
\\
 0 & 0 & 0 & 0 & 0 & 0 & 0 & 0 & 0 & 0 & 1 & 0 & 0 & 0 & 0 & 0 
\\
 1 & 0 & 0 & 0 & 0 & 0 & 0 & 0 & 0 & 0 & 0 & 0 & 0 & 0 & 0 & 0 
\end{array}\right)
\end{equation}
}
which has the following eigenvalues: 1 with multiplicity 7, -1 simple and 0 with multiplicity 0. Nevertheless, as in all previous cases, the matrix $A$ is diagonalizable and the asymptotic vector can be computed as a limit:
{\small
\begin{equation}
\left[\begin{array}{c}
\frac{v_{1}}{2}+\frac{v_{16}}{2}-\frac{\left(-v_{1}+v_{16}\right) \left(-1\right)^{n}}{2} \\ 0 \\ 0 \\ v_{4} \\ 0 \\ v_{2}+v_{5}+v_{6}+v_{8}+v_{14} \\ v_{7} \\ 0 \\ 0 \\ v_{10} \\ v_{3}+v_{9}+v_{11}+v_{12}+v_{15} \\ 0 \\ v_{13} \\ 0 \\ 0 \\ \frac{v_{1}}{2}+\frac{v_{16}}{2}+\frac{\left(-v_{1}+v_{16}\right) \left(-1\right)^{n}}{2} 
\end{array}\right]
\end{equation}
}
Note that periodic dynamics can appear for initial conditions with $v_1,v_{16}$ distinct from 0.  It is also interesting to remark that some state are never achieved for any initial condition. Specifically, components $v_2, v_3, v_5, v_8,v_9, v_{12}, v_{14}, v_{15}$  are null that corresponds with states $(0,0,0,1)$, $(0,0,1,0)$,  $(0,1,0,0)$, $(0,1,1,1)$, $(1,0,0,0)$, $(1,0,1,1)$, $(1,1,0,1)$, $(1,1,0,1)$, $(1,1,1,0)$, respectively.  % son las que tienen un unico uno o un unico cero.
Note too that $v_4, v_7, v_{10}$ and $v_{13}$ are pure absorbing states, corresponding respectively to: $(0,0,1,1)$, $(0,1,1,0)$, $(1,0,0,1)$, $(1,1,0,0)$. % dos unos y por tanto dos ceros seguidos. 

\subsection{Edge $r = 1$}
This edge represents the PCA where a non-standard contagion occurs in case of majority in the neighborhood. When a tie occurs, the probability of keeping the state of the central site depends on $p$. The Stochastic Transition Probability for the Markov Process that describes the dynamics is given by:

{\scriptsize
\begin{equation}
A  = 
\left(\begin{array}{cccccccccccccccc}
1 & 0 & 0 & 0 & 0 & 0 & 0 & 0 & 0 & 0 & 0 & 0 & 0 & 0 & 0 & 0 
\\
 p^{2} & 0 & p q  & 0 & 0 & 0 & 0 & 0 & p q  & 0 & q^{2} & 0 & 0 & 0 & 0 & 0 
\\
 p^{2} & p q  & 0 & 0 & p q  & q^{2} & 0 & 0 & 0 & 0 & 0 & 0 & 0 & 0 & 0 & 0 
\\
 p^{2} q^{2} & p^{3} q  & p^{3} q  & p^{4} & p \,q^{3} & p^{2} q^{2} & p^{2} q^{2} & p^{3} q  & p \,q^{3} & p^{2} q^{2} & p^{2} q^{2} & p^{3} q  & q^{4} & p \,q^{3} & p \,q^{3} & p^{2} q^{2} 
\\
 p^{2} & 0 & p q  & 0 & 0 & 0 & 0 & 0 & p q  & 0 & q^{2} & 0 & 0 & 0 & 0 & 0 
\\
 0 & 0 & 0 & 0 & 0 & 0 & 0 & 0 & 0 & 0 & 1 & 0 & 0 & 0 & 0 & 0 
\\
 p^{2} q^{2} & p \,q^{3} & p^{3} q  & p^{2} q^{2} & p^{3} q  & p^{2} q^{2} & p^{4} & p^{3} q  & p \,q^{3} & q^{4} & p^{2} q^{2} & p \,q^{3} & p^{2} q^{2} & p \,q^{3} & p^{3} q  & p^{2} q^{2} 
\\
 0 & 0 & 0 & 0 & 0 & 0 & 0 & 0 & 0 & 0 & q^{2} & p q  & 0 & 0 & p q  & p^{2} 
\\
 p^{2} & p q  & 0 & 0 & p q  & q^{2} & 0 & 0 & 0 & 0 & 0 & 0 & 0 & 0 & 0 & 0 
\\
 p^{2} q^{2} & p^{3} q  & p \,q^{3} & p^{2} q^{2} & p \,q^{3} & p^{2} q^{2} & q^{4} & p \,q^{3} & p^{3} q  & p^{4} & p^{2} q^{2} & p^{3} q  & p^{2} q^{2} & p^{3} q  & p \,q^{3} & p^{2} q^{2} 
\\
 0 & 0 & 0 & 0 & 0 & 1 & 0 & 0 & 0 & 0 & 0 & 0 & 0 & 0 & 0 & 0 
\\
 0 & 0 & 0 & 0 & 0 & q^{2} & 0 & p q  & 0 & 0 & 0 & 0 & 0 & p q  & 0 & p^{2} 
\\
 p^{2} q^{2} & p \,q^{3} & p \,q^{3} & q^{4} & p^{3} q  & p^{2} q^{2} & p^{2} q^{2} & p \,q^{3} & p^{3} q  & p^{2} q^{2} & p^{2} q^{2} & p \,q^{3} & p^{4} & p^{3} q  & p^{3} q  & p^{2} q^{2} 
\\
 0 & 0 & 0 & 0 & 0 & 0 & 0 & 0 & 0 & 0 & q^{2} & p q  & 0 & 0 & p q  & p^{2} 
\\
 0 & 0 & 0 & 0 & 0 & q^{2} & 0 & p q  & 0 & 0 & 0 & 0 & 0 & p q  & 0 & p^{2} 
\\
 0 & 0 & 0 & 0 & 0 & 0 & 0 & 0 & 0 & 0 & 0 & 0 & 0 & 0 & 0 & 1 
\end{array}\right)
\end{equation}
}
Still for this size, the eigenvalues of the matrix can be computed analytically yielding:
{\small
\begin{equation}
\lambda =\left(\begin{array}{c}
-2 p^{2}+2 p  
\\
 2 p^{2}-2 p  
\\
 -2 p^{2}+2 p  
\\
 2 p^{2}-2 p  
\\
 1 
\\
 1 
\\
 1 
\\
 -1 
\\
 4 p^{3}-6 p^{2}+4 p -1 
\\
 4 p^{3}-6 p^{2}+4 p -1 
\\
 4 p^{2}-4 p +1 
\\
 4 p^{4}-8 p^{3}+8 p^{2}-4 p +1 
\\
 0 
\\
 0 
\\
 0 
\\
 0 
\end{array}\right)
\end{equation}
}

As before, the existence of more than one eigenvalue with modulus 1 indicate the appearance of multistability and periodicity for particular initial conditions. 

We conclude this section considering the limit cases: $(r=1,p=0)$ and $(r=1,p=1)$ that conrrespond with the deterministic rules 178 and 232, respectivaly

{\bf Vertex $p=0$}.  The Stochastic Transition Matrix that defines the Markov Process in this case is:
{\small
\begin{equation}
\left(\begin{array}{cccccccccccccccc}
1 & 0 & 0 & 0 & 0 & 0 & 0 & 0 & 0 & 0 & 0 & 0 & 0 & 0 & 0 & 0 
\\
 0 & 0 & 0 & 0 & 0 & 0 & 0 & 0 & 0 & 0 & 1 & 0 & 0 & 0 & 0 & 0 
\\
 0 & 0 & 0 & 0 & 0 & 1 & 0 & 0 & 0 & 0 & 0 & 0 & 0 & 0 & 0 & 0 
\\
 0 & 0 & 0 & 0 & 0 & 0 & 0 & 0 & 0 & 0 & 0 & 0 & 1 & 0 & 0 & 0 
\\
 0 & 0 & 0 & 0 & 0 & 0 & 0 & 0 & 0 & 0 & 1 & 0 & 0 & 0 & 0 & 0 
\\
 0 & 0 & 0 & 0 & 0 & 0 & 0 & 0 & 0 & 0 & 1 & 0 & 0 & 0 & 0 & 0 
\\
 0 & 0 & 0 & 0 & 0 & 0 & 0 & 0 & 0 & 1 & 0 & 0 & 0 & 0 & 0 & 0 
\\
 0 & 0 & 0 & 0 & 0 & 0 & 0 & 0 & 0 & 0 & 1 & 0 & 0 & 0 & 0 & 0 
\\
 0 & 0 & 0 & 0 & 0 & 1 & 0 & 0 & 0 & 0 & 0 & 0 & 0 & 0 & 0 & 0 
\\
 0 & 0 & 0 & 0 & 0 & 0 & 1 & 0 & 0 & 0 & 0 & 0 & 0 & 0 & 0 & 0 
\\
 0 & 0 & 0 & 0 & 0 & 1 & 0 & 0 & 0 & 0 & 0 & 0 & 0 & 0 & 0 & 0 
\\
 0 & 0 & 0 & 0 & 0 & 1 & 0 & 0 & 0 & 0 & 0 & 0 & 0 & 0 & 0 & 0 
\\
 0 & 0 & 0 & 1 & 0 & 0 & 0 & 0 & 0 & 0 & 0 & 0 & 0 & 0 & 0 & 0 
\\
 0 & 0 & 0 & 0 & 0 & 0 & 0 & 0 & 0 & 0 & 1 & 0 & 0 & 0 & 0 & 0 
\\
 0 & 0 & 0 & 0 & 0 & 1 & 0 & 0 & 0 & 0 & 0 & 0 & 0 & 0 & 0 & 0 
\\
 0 & 0 & 0 & 0 & 0 & 0 & 0 & 0 & 0 & 0 & 0 & 0 & 0 & 0 & 0 & 1 
\end{array}\right)
\end{equation}
}
Its eigenvalues are: 1 with multiplicity 5, -1 with multiplicity 3 and 0 with multiplicity 8. 

The asymptotic vector is then given by:

{\small
\begin{equation}
\left[\begin{array}{c}
v_{1} \\ 0 \\ 0 \\ \frac{v_{4}}{2}+\frac{v_{13}}{2}-\frac{\left(-v_{4}+v_{13}\right) \left(-1\right)^{n}}{2} \\ 0 \\ \frac{v_{2}}{2}+\frac{v_{3}}{2}+\frac{v_{5}}{2}+\frac{v_{6}}{2}+\frac{v_{8}}{2}+\frac{v_{9}}{2}+\frac{v_{11}}{2}+\frac{v_{12}}{2}+\frac{v_{14}}{2}+\frac{v_{15}}{2}-
\\
\frac{\left(-v_{2}+v_{3}-v_{5}-v_{6}-v_{8}+v_{9}+v_{11}+v_{12}-v_{14}+v_{15}\right) \left(-1\right)^{n}}{2} \\ \frac{v_{7}}{2}+\frac{v_{10}}{2}-\frac{\left(-v_{7}+v_{10}\right) \left(-1\right)^{n}}{2} \\ 0 \\ 0 \\ \frac{v_{7}}{2}+\frac{v_{10}}{2}+\frac{\left(-v_{7}+v_{10}\right) \left(-1\right)^{n}}{2} \\ \frac{v_{2}}{2}+\frac{v_{3}}{2}+\frac{v_{5}}{2}+\frac{v_{6}}{2}+\frac{v_{8}}{2}+\frac{v_{9}}{2}+\frac{v_{11}}{2}+\frac{v_{12}}{2}+\frac{v_{14}}{2}+\frac{v_{15}}{2}+
\\
\frac{\left(-v_{2}+v_{3}-v_{5}-v_{6}-v_{8}+v_{9}+v_{11}+v_{12}-v_{14}+v_{15}\right) \left(-1\right)^{n}}{2} \\ 0 \\ \frac{v_{4}}{2}+\frac{v_{13}}{2}+\frac{\left(-v_{4}+v_{13}\right) \left(-1\right)^{n}}{2} \\ 0 \\ 0 \\ v_{16} 
\end{array}\right]
\end{equation}
}

{\bf Vertex $p=1$}. The other vertex in the edge $r=1$ corresponds with the deterministic rule 232. The Stochastic Transition Matrix  is:
{\small
\begin{equation}
\left(\begin{array}{cccccccccccccccc}
1 & 0 & 0 & 0 & 0 & 0 & 0 & 0 & 0 & 0 & 0 & 0 & 0 & 0 & 0 & 0 
\\
 1 & 0 & 0 & 0 & 0 & 0 & 0 & 0 & 0 & 0 & 0 & 0 & 0 & 0 & 0 & 0 
\\
 1 & 0 & 0 & 0 & 0 & 0 & 0 & 0 & 0 & 0 & 0 & 0 & 0 & 0 & 0 & 0 
\\
 0 & 0 & 0 & 1 & 0 & 0 & 0 & 0 & 0 & 0 & 0 & 0 & 0 & 0 & 0 & 0 
\\
 1 & 0 & 0 & 0 & 0 & 0 & 0 & 0 & 0 & 0 & 0 & 0 & 0 & 0 & 0 & 0 
\\
 0 & 0 & 0 & 0 & 0 & 0 & 0 & 0 & 0 & 0 & 1 & 0 & 0 & 0 & 0 & 0 
\\
 0 & 0 & 0 & 0 & 0 & 0 & 1 & 0 & 0 & 0 & 0 & 0 & 0 & 0 & 0 & 0 
\\
 0 & 0 & 0 & 0 & 0 & 0 & 0 & 0 & 0 & 0 & 0 & 0 & 0 & 0 & 0 & 1 
\\
 1 & 0 & 0 & 0 & 0 & 0 & 0 & 0 & 0 & 0 & 0 & 0 & 0 & 0 & 0 & 0 
\\
 0 & 0 & 0 & 0 & 0 & 0 & 0 & 0 & 0 & 1 & 0 & 0 & 0 & 0 & 0 & 0 
\\
 0 & 0 & 0 & 0 & 0 & 1 & 0 & 0 & 0 & 0 & 0 & 0 & 0 & 0 & 0 & 0 
\\
 0 & 0 & 0 & 0 & 0 & 0 & 0 & 0 & 0 & 0 & 0 & 0 & 0 & 0 & 0 & 1 
\\
 0 & 0 & 0 & 0 & 0 & 0 & 0 & 0 & 0 & 0 & 0 & 0 & 1 & 0 & 0 & 0 
\\
 0 & 0 & 0 & 0 & 0 & 0 & 0 & 0 & 0 & 0 & 0 & 0 & 0 & 0 & 0 & 1 
\\
 0 & 0 & 0 & 0 & 0 & 0 & 0 & 0 & 0 & 0 & 0 & 0 & 0 & 0 & 0 & 1 
\\
 0 & 0 & 0 & 0 & 0 & 0 & 0 & 0 & 0 & 0 & 0 & 0 & 0 & 0 & 0 & 1 
\end{array}\right)
\end{equation}
}
Its eigenvalues are: 1 with multiplicity 7, -1 simple and 0 with multiplicity 8.

The asymptotic vector depends on the initial vector $v=(v_1,v_2,\ldots,v_{15},v_{16})$ as follows:
{\small
\begin{equation}
\left(\begin{array}{c}
v_{1}+v_{2}+v_{3}+v_{5}+v_{9} \\ 0 \\ 0 \\ v_{4} \\  0 \\  \frac{v_{6}}{2}+\frac{v_{11}}{2}-\frac{\left(-v_{6}+v_{11}\right) \left(-1\right)^{n}}{2} \\ v_{7} \\ 0 \\ 0 \\ v_{10} \\ \frac{v_{6}}{2}+\frac{v_{11}}{2}+\frac{\left(-v_{6}+v_{11}\right) \left(-1\right)^{n}}{2} \\  0 \\ v_{13} \\ 0 \\ 0 \\ v_{8}+v_{12}+v_{14}+v_{15}+v_{16} 
\end{array}\right)
\end{equation}
}

\subsection{Edge $p = 0$}

This edge corresponds to probabilistic rules with a certainty of keeping the current state of the central site in case of tie in its neighborhood. The Stoachastic Probabilistic Matrix of the associated Marvok Process is:

{\scriptsize
\begin{equation}
A  = 
\left(\begin{array}{cccccccccccccccc}
r^{4} & r^{3} s  & r^{3} s  & r^{2} s^{2} & r^{3} s  & r^{2} s^{2} & r^{2} s^{2} & r \,s^{3} & r^{3} s  & r^{2} s^{2} & r^{2} s^{2} & r \,s^{3} & r^{2} s^{2} & r \,s^{3} & r \,s^{3} & s^{4} 
\\
 0 & 0 & 0 & 0 & 0 & 0 & 0 & 0 & 0 & 0 & r^{2} & r s  & 0 & 0 & r s  & s^{2} 
\\
 0 & 0 & 0 & 0 & 0 & r^{2} & 0 & r s  & 0 & 0 & 0 & 0 & 0 & r s  & 0 & s^{2} 
\\
 0 & 0 & 0 & 0 & 0 & 0 & 0 & 0 & 0 & 0 & 0 & 0 & 1 & 0 & 0 & 0 
\\
 0 & 0 & 0 & 0 & 0 & 0 & 0 & 0 & 0 & 0 & r^{2} & r s  & 0 & 0 & r s  & s^{2} 
\\
 r^{2} s^{2} & r \,s^{3} & r^{3} s  & r^{2} s^{2} & r \,s^{3} & s^{4} & r^{2} s^{2} & r \,s^{3} & r^{3} s  & r^{2} s^{2} & r^{4} & r^{3} s  & r^{2} s^{2} & r \,s^{3} & r^{3} s  & r^{2} s^{2} 
\\
 0 & 0 & 0 & 0 & 0 & 0 & 0 & 0 & 0 & 1 & 0 & 0 & 0 & 0 & 0 & 0 
\\
 s^{2} & 0 & r s  & 0 & 0 & 0 & 0 & 0 & r s  & 0 & r^{2} & 0 & 0 & 0 & 0 & 0 
\\
 0 & 0 & 0 & 0 & 0 & r^{2} & 0 & r s  & 0 & 0 & 0 & 0 & 0 & r s  & 0 & s^{2} 
\\
 0 & 0 & 0 & 0 & 0 & 0 & 1 & 0 & 0 & 0 & 0 & 0 & 0 & 0 & 0 & 0 
\\
 r^{2} s^{2} & r^{3} s  & r \,s^{3} & r^{2} s^{2} & r^{3} s  & r^{4} & r^{2} s^{2} & r^{3} s  & r \,s^{3} & r^{2} s^{2} & s^{4} & r \,s^{3} & r^{2} s^{2} & r^{3} s  & r \,s^{3} & r^{2} s^{2} 
\\
 s^{2} & r s  & 0 & 0 & r s  & r^{2} & 0 & 0 & 0 & 0 & 0 & 0 & 0 & 0 & 0 & 0 
\\
 0 & 0 & 0 & 1 & 0 & 0 & 0 & 0 & 0 & 0 & 0 & 0 & 0 & 0 & 0 & 0 
\\
 s^{2} & 0 & r s  & 0 & 0 & 0 & 0 & 0 & r s  & 0 & r^{2} & 0 & 0 & 0 & 0 & 0 
\\
 s^{2} & r s  & 0 & 0 & r s  & r^{2} & 0 & 0 & 0 & 0 & 0 & 0 & 0 & 0 & 0 & 0 
\\
 s^{4} & r \,s^{3} & r \,s^{3} & r^{2} s^{2} & r \,s^{3} & r^{2} s^{2} & r^{2} s^{2} & r^{3} s  & r \,s^{3} & r^{2} s^{2} & r^{2} s^{2} & r^{3} s  & r^{2} s^{2} & r^{3} s  & r^{3} s  & r^{4} 
\end{array}\right)
\end{equation}
}
Again, matrix $A$ has multiple eigenvalues with modulus 1, specifically  
{\small
\begin{equation}
\lambda = \left(\begin{array}{c}
1 
\\
 -1 
\\
 1 
\\
 -1 
\\
 0 
\\
 0 
\\
 0 
\\
 0 
\\
 4 r^{2}-4 r +1 
\\
 -2 r^{3}+4 r^{2}-3 r +\frac{1}{2}+
\\
\frac{\sqrt{16 r^{6}-64 r^{5}+80 r^{4}-40 r^{3}+12 r^{2}-4 r +1}}{2} 
\\
 -2 r^{3}+4 r^{2}-3 r +\frac{1}{2}-
\\
\frac{\sqrt{16 r^{6}-64 r^{5}+80 r^{4}-40 r^{3}+12 r^{2}-4 r +1}}{2} 
\\
 2 r^{3}-2 r^{2}+r -\frac{1}{2}+\frac{\sqrt{16 r^{6}-32 r^{5}+40 r^{3}-28 r^{2}+4 r +1}}{2} 
\\
 2 r^{3}-2 r^{2}+r -\frac{1}{2}-\frac{\sqrt{16 r^{6}-32 r^{5}+40 r^{3}-28 r^{2}+4 r +1}}{2} 
\\
 2 r^{4}-4 r^{3}+3 r^{2}-r +\frac{1}{2}+
\\
\frac{\sqrt{16 r^{8}-64 r^{7}+80 r^{6}-16 r^{5}-52 r^{4}+56 r^{3}-24 r^{2}+4 r +1}}{2} 
\\
 2 r^{4}-4 r^{3}+3 r^{2}-r +\frac{1}{2}-
\\
\frac{\sqrt{16 r^{8}-64 r^{7}+80 r^{6}-16 r^{5}-52 r^{4}+56 r^{3}-24 r^{2}+4 r +1}}{2} 
\\
 -2 r^{2}+2 r  
\end{array}\right)
\end{equation}
}
The existence of two eigenvalues with modulus 1, 1 and -1, both with multiplicity 2 implies multistability and, consequently, a dependence of the asymptotic dynamics on the initial vector.

\subsection{Edege $p = 1$}

The opposite edge means that, in case of tie in the neighborhood, the centra site changes its current state with probability 1. In this case, the probability of adoption of the states of the neighbors when a majority occurs depends on the $r$-value. This dependence is reflected in the corresponding Stochastic Transition Matrix:
{\scriptsize
\begin{equation}
A=\left(\begin{array}{cccccccccccccccc}
r^{4} & r^{3} s  & r^{3} s  & r^{2} s^{2} & r^{3} s  & r^{2} s^{2} & r^{2} s^{2} & r \,s^{3} & r^{3} s  & r^{2} s^{2} & r^{2} s^{2} & r \,s^{3} & r^{2} s^{2} & r \,s^{3} & r \,s^{3} & s^{4} 
\\
 r^{2} & r s  & 0 & 0 & r s  & s^{2} & 0 & 0 & 0 & 0 & 0 & 0 & 0 & 0 & 0 & 0 
\\
 r^{2} & 0 & r s  & 0 & 0 & 0 & 0 & 0 & r s  & 0 & s^{2} & 0 & 0 & 0 & 0 & 0 
\\
 0 & 0 & 0 & 1 & 0 & 0 & 0 & 0 & 0 & 0 & 0 & 0 & 0 & 0 & 0 & 0 
\\
 r^{2} & r s  & 0 & 0 & r s  & s^{2} & 0 & 0 & 0 & 0 & 0 & 0 & 0 & 0 & 0 & 0 
\\
 r^{2} s^{2} & r \,s^{3} & r^{3} s  & r^{2} s^{2} & r \,s^{3} & s^{4} & r^{2} s^{2} & r \,s^{3} & r^{3} s  & r^{2} s^{2} & r^{4} & r^{3} s  & r^{2} s^{2} & r \,s^{3} & r^{3} s  & r^{2} s^{2} 
\\
 0 & 0 & 0 & 0 & 0 & 0 & 1 & 0 & 0 & 0 & 0 & 0 & 0 & 0 & 0 & 0 
\\
 0 & 0 & 0 & 0 & 0 & s^{2} & 0 & r s  & 0 & 0 & 0 & 0 & 0 & r s  & 0 & r^{2} 
\\
 r^{2} & 0 & r s  & 0 & 0 & 0 & 0 & 0 & r s  & 0 & s^{2} & 0 & 0 & 0 & 0 & 0 
\\
 0 & 0 & 0 & 0 & 0 & 0 & 0 & 0 & 0 & 1 & 0 & 0 & 0 & 0 & 0 & 0 
\\
 r^{2} s^{2} & r^{3} s  & r \,s^{3} & r^{2} s^{2} & r^{3} s  & r^{4} & r^{2} s^{2} & r^{3} s  & r \,s^{3} & r^{2} s^{2} & s^{4} & r \,s^{3} & r^{2} s^{2} & r^{3} s  & r \,s^{3} & r^{2} s^{2} 
\\
 0 & 0 & 0 & 0 & 0 & 0 & 0 & 0 & 0 & 0 & s^{2} & r s  & 0 & 0 & r s  & r^{2} 
\\
 0 & 0 & 0 & 0 & 0 & 0 & 0 & 0 & 0 & 0 & 0 & 0 & 1 & 0 & 0 & 0 
\\
 0 & 0 & 0 & 0 & 0 & s^{2} & 0 & r s  & 0 & 0 & 0 & 0 & 0 & r s  & 0 & r^{2} 
\\
 0 & 0 & 0 & 0 & 0 & 0 & 0 & 0 & 0 & 0 & s^{2} & r s  & 0 & 0 & r s  & r^{2} 
\\
 s^{4} & r \,s^{3} & r \,s^{3} & r^{2} s^{2} & r \,s^{3} & r^{2} s^{2} & r^{2} s^{2} & r^{3} s  & r \,s^{3} & r^{2} s^{2} & r^{2} s^{2} & r^{3} s  & r^{2} s^{2} & r^{3} s  & r^{3} s  & r^{4} 
\end{array}\right)
\end{equation}
}
whose eigenvalues are:
{\scriptsize
\begin{equation}
\lambda=\left(\begin{array}{c}
1 
\\
 1 
\\
 1 
\\
 1 
\\
 0 
\\
 0 
\\
 0 
\\
 0 
\\
 4 r^{2}-4 r +1 
\\
 -2 r^{2}+2 r  
\\
 2 r^{4}-4 r^{3}+3 r^{2}-r +\frac{1}{2}+
\\
\frac{\sqrt{16 r^{8}-64 r^{7}+80 r^{6}-16 r^{5}-52 r^{4}+56 r^{3}-24 r^{2}+4 r +1}}{2} 
\\
 2 r^{4}-4 r^{3}+3 r^{2}-r +\frac{1}{2}-
\\
\frac{\sqrt{16 r^{8}-64 r^{7}+80 r^{6}-16 r^{5}-52 r^{4}+56 r^{3}-24 r^{2}+4 r +1}}{2} 
\\
 -2 r^{3}+2 r^{2}-r +\frac{1}{2}+\frac{\sqrt{16 r^{6}-32 r^{5}+40 r^{3}-28 r^{2}+4 r +1}}{2} 
\\
 -2 r^{3}+2 r^{2}-r +\frac{1}{2}-\frac{\sqrt{16 r^{6}-32 r^{5}+40 r^{3}-28 r^{2}+4 r +1}}{2} 
\\
 2 r^{3}-4 r^{2}+3 r -\frac{1}{2}+
\\
\frac{\sqrt{16 r^{6}-64 r^{5}+80 r^{4}-40 r^{3}+12 r^{2}-4 r +1}}{2} 
\\
 2 r^{3}-4 r^{2}+3 r -\frac{1}{2}-
\\
\frac{\sqrt{16 r^{6}-64 r^{5}+80 r^{4}-40 r^{3}+12 r^{2}-4 r +1}}{2} 
\end{array}\right)
\end{equation}
}

\section{Concluding remarks}\label{concluding}

We have presented a Probabilistic Cellular Automaton (PCA) that generates a parametric square defined by $0 \leq p, r \leq 1$, whose vertices correspond to the deterministic Wolfram rules 23 $(p=0, r=0)$, 77 $(p=1, r=0)$, 178 $(p=0, r=1)$, and 232 $(p=1, r=1)$.  This PCA can be applied to study opinion dynamics in situations where agents exhibit hesitation, where $p$ determines the likelihood of an agent keeping its current state when its neighborhood is evenly split, and $r$ quantifies the agent’s reluctance to adopt a position that goes against the majority. By combining these two forms of uncertainty, the model provides a framework for exploring how hesitation influences the evolution of opinions within social systems.

Low size PCA can be analyzed using a Markov process description, where the states are given by the coordinates of the stochastic vector, which evolves according to the corresponding Stochastic Transition Matrix (STM). For small system sizes, the asymptotic behavior can be determined as a function of the initial conditions and the parameters $p$, $r$. We have shown that, for $0 < p, r < 1$, the STM is diagonalizable and positive. This implies that it has a simple eigenvalue equal to 1, while all other eigenvalues have modulus less than 1. Therefore, the asymptotic dynamics of the PCA are determined by the eigenvector associated with the eigenvalue 1. Consequently, a unique stationary stochastic vector is reached for all initial conditions. In contrast, the positivity of the STM is lost along the edges of the square, i.e., when either $p=0$ or $p=1$ or $r=0$ or $r=1$, and in particular at its vertices. At these points, the asymptotic vector depends on both the initial conditions and the parameter values.

It is worth noting that the asymptotic behavior of the PCA is not continuous along some the edges of the square, nor at its vertices. These dynamical discontinuities arise from the appearance of additional eigenvalues with modulus 1 along the edges. In particular, it has been shown that periodic behavior (of period two) that does not occur in the interior of the square, may arise on the boundary, and especially at the vertices. Moreover, these conditions, namely, the probability distributions over configurations that lead to periodic behavior, can be explicitly identified.

We conclude by noting that the analysis presented here can be straightforwardly extended to PCA of larger size. However, the computational cost increases rapidly, since a system of size $L$ leads to a Markov process of dimension $2^L$, which becomes prohibitively large even for moderate values of $L$.

% -------------------------------------------------
% BIBLIOGRAPHY
% -------------------------------------------------

%\bibliography{cas-refs}

\end{document}